\documentclass[useAMS,referee,usenatbib]{biom}
\pdfoutput = 1
\usepackage{caption}
\usepackage{amsmath}
\usepackage[figuresright]{rotating}

\usepackage{bm}
\usepackage{natbib}

\usepackage{booktabs}
\usepackage{amsfonts}
\usepackage{multirow}
\usepackage{arydshln}

\def\E{\mathbb{E}}
\def\P{\mathbb{P}}
\def\given{\, | \,}

\allowdisplaybreaks

\usepackage{xcolor}

\title[Estimating Time-Varying Direct and Indirect Causal Excursion Effects with Longitudinal Binary Outcomes]{Estimating Time-Varying Direct and Indirect \\
Causal Excursion Effects with Longitudinal Binary Outcomes}

\author{Jieru Shi$^*$\email{herashi@umich.edu}, 
Zhenke Wu$^{**}$\email{zhenkewu@umich.edu}, and 
Walter Dempsey$^{***}$\email{wdem@umich.edu} \\
Department of Biostatistics, University of Michigan, Ann Arbor, MI, USA}

\begin{document}


\date{{\it Received October} 2007. {\it Revised February} 2008.  {\it
Accepted March} 2008.}



\pagerange{\pageref{firstpage}--\pageref{lastpage}} 
\volume{64}
\pubyear{2008}
\artmonth{December}


\doi{10.1111/j.1541-0420.2005.00454.x}


\label{firstpage}


\begin{abstract}
Construction of just-in-time adaptive interventions, such as prompts delivered by mobile apps to promote and maintain behavioral change, requires knowledge about time-varying moderated effects to inform when and how we deliver intervention options. Micro-randomized trials (MRT) have emerged as a sequentially randomized design to gather requisite data for effect estimation. The existing literature \citep{qian2020estimating,boruvka2018assessing,dempsey2020stratified} has defined a general class of causal estimands, referred to as ``causal excursion effects", to assess the time-varying moderated effect. However, there is limited statistical literature on how to address potential between-cluster treatment effect heterogeneity and within-cluster interference in a sequential treatment setting for longitudinal binary outcomes. In this paper, based on a cluster conceptualization of the potential outcomes, we define a larger class of direct and indirect causal excursion effects for proximal and lagged binary outcomes, and propose a new inferential procedure that addresses effect heterogeneity and interference. We provide theoretical guarantees of consistency and asymptotic normality of the estimator. Extensive simulation studies confirm our theory empirically and show the proposed procedure provides consistent point estimator and interval estimates with valid coverage. Finally, we analyze a data set from a multi-institution MRT study to assess the time-varying moderated effects of mobile prompts upon binary study engagement outcomes. 

\end{abstract}
%


\maketitle


%

\section{Introduction}
\label{s:intro}

As internet-enabled devices become increasingly popular, mobile devices such as smartphones and wearable devices are used to sense the current context of the individual, as well as to deliver interventions intended to promote healthy behaviors and health-related behavioral change \citep[e.g.,][]{free2013effectiveness}. 
Examples include medication reminders, motivational messages, and reminders to engage in physical activity. 

There is intense need in gathering sequential experimental data to guide the development of just-in-time adaptive interventions (JITAIs) \citep{nahum2018just}. JITAIs are mHealth technologies that aim to deliver the right intervention components at the right times to offer optimal support to promote individuals’ healthy behaviors. Micro-randomized trials (MRTs) \citep{klasnja2015} have emerged as a popular sequentially randomized design where each subject is repeatedly randomized to receive one of the multiple treatment options, often at hundreds or even thousands of times, over the course of the trial. In all cases the randomization probabilities are determined as part of the design and thus known. Between randomizations, the individual's current or recent context is collected via sensors and/or self-reports, and after each randomization a proximal, near-time outcome is collected. The time-varying treatments, information collected between treatments, and proximal outcomes constitute longitudinal data for use in assessing whether the treatment has an effect on the proximal outcome at each treatment occasion. For the domain scientists, estimates of these treatment effects are crucial to inform the development of mobile health interventions.

Many MRTs to date have longitudinal binary outcomes as the primary outcome. For example, in an MRT conducted by JOOL Health \citep{2018prompt}, the researchers intended to test whether sending a push notification containing a contextually tailored health message versus not sending push notification results in an increased likelihood of proximal engagement with the app. The outcome in this case is user engagement with the app, which is a binary variable. Similarly, the Substance Abuse Research Assistance (SARA) study \citep{rabbi2018toward} has a proximal outcome of whether participants completed the survey and active tasks. In BariFit \citep{Barifit}, the proximal outcome for the food tracking reminder is whether the participant completes their food log on that day or not. 
In the following, we present an example that highlights the limitations of existing work and the need for novel estimands and estimation techniques. 


\subsection{Motivating Example}
\label{sec:motivatingexample}
This work is motivated by our involvement in the Intern Health Study \citep[IHS;][]{Necamp2020}, which is a 6-month micro-randomized trial on 1,565 medical interns. Medical internship during the first year of physician residency training is highly stressful,  resulting in depression rates several folds higher than those of the general population. In order to understand whether mobile interventions improve participants' mental health, enrolled medical interns were randomized weekly to receive either mood, activity, or sleep notifications or receive no notifications for that week (probability 1/4 each). During a week when notifications are provided, interns were randomized daily with 50\% probability to receive a notification. Mood valence surveys were administered daily to all interns. Participants' engagement with the mobile app is essential. We are interested in whether targeted notifications affect participants' daily survey completion. In this context, the proximal outcome is whether or not a participant completes the daily mood survey, which is a binary longitudinal outcome. 





\begin{figure}
 \centerline{\includegraphics[width=4in]{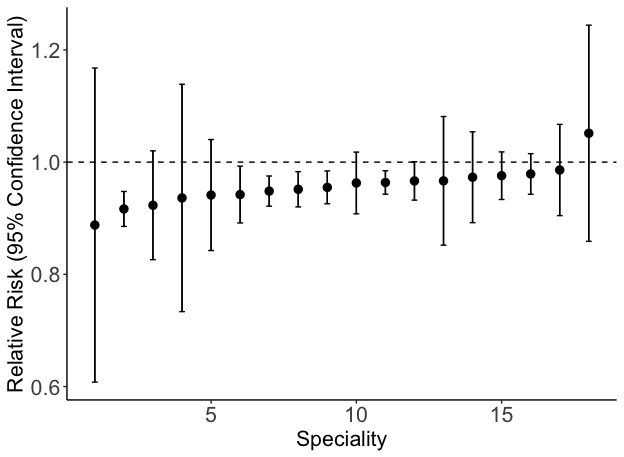}}
\caption{Observed treatment effect heterogeneity of the estimated relative risks using the existing EMEE across the clusters (specialties with greater than or equal to 6 people).}
\label{fig:clusterdif}
\end{figure}

\citet{qian2020estimating} proposed an estimator of the marginal excursion effect (EMEE) to examine whether a particular time-varying intervention has an effect on a longitudinal binary outcome. This inferential method for data collected in an MRT relies on two key assumptions: no between-individual interference and no treatment effect heterogeneity. However, the 1,565 interns in the IHS represented 321 residency institutions and 42 specialties, resulting in naturally clustered subjects. In other words, we are studying an MRT in which there is a pre-defined clustering structure, with randomization still at the individual level. An exploratory analysis is conducted to illustrate this issue. The clusters are defined by specialties, in particular, we remove specialties with fewer than six participants, and apply EMEE to every cluster separately to estimate the treatment effects. Figure \ref{fig:clusterdif} presents specialty-specific effect estimates of daily prompts upon daily survey completion, with evident 
cluster-level treatment effect heterogeneity. Motivated by this example, we are interested in performing inferences on novel causal excursion effects when cluster structures exist.

\subsection{Main Contribution}
For proximal and lagged binary outcomes, we propose a family of estimands of causal excursion effects to account for potential between-cluster treatment effect heterogeneity and within-cluster interference. We define direct and pairwise indirect causal excursion effects accounting for cluster structure. In particular, the direct effect characterizes the effect of an intervention when a person receives treatment versus not; the pairwise indirect effect accounts for how another individual's treatment status within the same cluster affects one's proximal or lagged outcome. 

In this paper, we provide conditions that ensure identification and propose a new estimation method termed ``cluster-based estimator of the marginal excursion effect'' (C-EMEE) to consistently estimate direct and pairwise indirect causal excursion effects with longitudinal binary outcomes. A robust variance estimator is used to ensure consistent estimation of asymptotic variance. A small sample correction \citep{mancl2001covariance} is added when the sample size is under 50. Theoretical results, including consistency and asymptotic normality of the proposed estimators, are presented. From a practical perspective, our proposed methods for estimating time-varying causal excursion effects include prior methods such as EMEE as a special case, and are more well-suited for real-world environments in which clustering structures exist. 

The rest of the paper is organized as follows.
We begin with a brief literature review in Section \ref{sec:Preliminaries}. In Section \ref{section:prox_effects_pot_outcome}, we define novel direct and pairwise indirect causal excursion effects for longitudinal binary outcomes. Following this, in Section \ref{section:estimation}, the cluster-based estimators of the marginal excursion effect are presented. The asymptotic properties of these estimators are assessed both theoretically and numerically using a variety of simulation scenarios in Section \ref{section:sims}. In Section \ref{sec:casestudy}, we apply the proposed methods to data from the Intern Health Study \citep{Necamp2020}, a mobile intervention study conducted to investigate whether mobile prompt notifications can be used to impact participants' engagement on completing their daily self-reported mood surveys. The paper concludes with a brief summary and discussion of future directions.

\section{Preliminaries}
\label{sec:Preliminaries}

\subsection{Micro-Randomized Trials (MRT)}
An MRT consists of a sequence of within-subject decision times $t=1,\ldots,T$ at which treatment options may be randomly assigned.  The resulting longitudinal data for any subject can be summarized as
$$
\{ O_0, O_1, A_1, O_2, A_2, \ldots, O_T, A_T, O_{T+1} \},
$$
where $O_0$ is the baseline information, $O_t$ is the information collected between time $t-1$ and $t$ which may include mobile sensor and wearable data, and $A_t$ is the treatment option provided at time $t$. For simplicity we assume that there are two treatment options: $A_t \in \{0,1\}$. We use over-bar to denote a sequence of variables up to a time point; for example $\bar A_t =(A_1,\ldots,A_t)$. Information accrued up to time $t$ is represented by the history $H_t := \{ O_0, O_1, A_1, \ldots, A_{t-1}, O_t \} = \{\bar O_t,\bar A_{t-1}\}$.

In an MRT, $A_t$ results from realized randomization at decision occasion $t$; here, we assume this randomization probability depends on the complete observed history $H_t$ and is denoted by $p_t (A_t \given H_t)$ and $\mathbf{p} = \{p_t (A_t \given H_t)\}_{t=1}^T$. Treatment options are designed to impact a proximal or possibly lagged response $Y_{t,\Delta}$ which is a known function of the participant's data within a subsequent window of length $\Delta \geq 1$:  $Y_{t,\Delta}=y(H_{t+\Delta})$; when $\Delta=1$, we set $Y_{t,1}=y(H_{t+1})$ \citep{dempsey2020stratified}. In this paper, $Y_{t,\Delta}$ is a binary random variable. 

\begin{remark}[Availability]
At some decision points, it is inappropriate for scientific, ethical, or burden reasons to provide treatment. To address this, the observation vector $O_t$ also contains an indicator of availability: $I_t = 1$ if available for treatment and $I_t = 0$ otherwise. Availability at time $t$ is determined before treatment randomization. When an individual is unavailable for treatment ($I_t = 0$), then no treatment is provided ($A_t = 0$). In this paper, for simplicity, individuals are assumed always available for treatment. All methods can be readily extended to account for availability as in \cite{boruvka2018assessing} and \cite{qian2020estimating}. 
\end{remark}

\subsection{Existing Estimand and Inferential Method}
\label{section:standardmrtmethods}


Substantial work has been done by \citet{boruvka2018assessing}, \citet{dempsey2020stratified} and \citet{shi2021} on the estimation of causal excursion effects with continuous outcomes. Later,  \cite{qian2020estimating} addressed the unique challenges in the binary outcome setting by adopting a log relative risk model for the causal excursion effect of $A_t$ on $Y_{t,\Delta}$. Assuming availability at all time points, we have \textit{causal excursion effects} for binary outcome defined by:
\begin{align}
        \beta_{\mathbf{p},\pi,\Delta} (t;s) &= \log \frac{\E \left[ Y_{t,\Delta}(\bar A_{t-1}, 1,\tilde{A}_{t+1:(t+\Delta -1)})\given S_t(\bar A_{t-1})=s \right]}{\E \left[ Y_{t,\Delta}(\bar A_{t-1}, 0,\tilde{A}_{t+1:(t+\Delta -1)})\given S_t(\bar A_{t-1})=s \right]}. 
\end{align}


$S_t$ is a time-varying potential effect moderator (or ``state") and a deterministic function of the observed history $H_t$. Here, we assume the reference distribution for treatment assignments from $t+1$ to $t+\Delta-1$ ($\Delta>1$) is given by a randomization probability generically represented by~$\pi_{u}(a_{u} | H_{u}), u=t+1,\ldots, t+\Delta-1$ and let $\pi=\{\pi_{u}\}_{u=t+1}^{t+\Delta-1}$.  This generalization contains previous definitions such as lagged effects~\citep{boruvka2018assessing} where $\pi_{u} = p_{u}$ and deterministic choices such as $a_{t+1:(t+\Delta-1)} = {\bf 0}$~\citep{dempsey2020stratified,qian2020estimating}, where $\pi_{u} = {\bf 1}\{a_{u} = 0\}$ and ${\bf 1}\{\cdot\}$ is the indicator function. 

Under the assumption that $\beta_{\mathbf{p},\pi,\Delta} (t;s) = f_t(s)^\top \beta^\star$, where $f_t(S_t) \in \mathbb{R}^q$ is a feature vector depending only on state $S_t = s$ and decision point $t$, a consistent estimator of $\beta^\star$ can be obtained by applying the EMEE method \citep{qian2020estimating}:
\begin{equation}
\label{eq:mrtstandard}
\mathbb{P}_n \left[ \sum_{t=1}^{T-\Delta +1}  W_{t,\Delta} W_t e^{-A_{t} f_t (S_t)^\top \beta} \left( Y_{t,\Delta} - e^{g_t(H_t)^\top \alpha + A_t f_t (S_t)^\top \beta} \right) \left( \begin{array}{c} g_t(H_t) \\ (A_{t} - \tilde p_t (A_{t}=1 | S_t) ) f_t(S_t) \end{array} \right) \right] = 0,
\end{equation}
where~$\mathbb{P}_n$ is defined as the average of a function over the sample, and $g_t(H_t) \in \mathcal{R}^p$ is a vector of control variables chosen to help reduce the variance of the estimators and construct more powerful test statistics. The product $W_{t,\Delta}=\prod_{u=t+1}^{t+\Delta-1}\pi_u(A_u \given H_u)/p_u(A_u \given H_u)$ is the standard inverse probability weighting for settings with $\Delta >1$, and $W_{t,\Delta}=1$ when $\Delta=1$. $W_t(H_t) = \tilde p (A_t | S_t) / p_t (A_t | H_t)$ is a weight where the numerator is an arbitrary function with range $(0,1)$ that only depends on $S_t$ and the denominator is the MRT specified randomization probability.

\begin{remark}
Typically, lagged effects of interest have short time horizons, therefore $\Delta$ is chosen to be small in order to provide more detailed insight into the causal excursion effect. Increasing $\Delta$ will result in exponential growth of variance since the weight $W_{t,\Delta}$ is a product of $\Delta-1$ density ratios. This is known as the curse of the horizon \citep{liu2020}.
\end{remark}

\section{Cluster-level proximal treatment effects}
\label{section:prox_effects_pot_outcome}

The primary question of interest is whether treatment has an immediate and/or lagged effect on the binary outcome. Due to potential within cluster interference, we consider two types of moderated effects: a direct and a pairwise indirect effect. 


Consider a cluster of size $G$. We have $\bar a_{t,j} = (a_{1,j},\ldots, a_{t,j})$ denote the sequence of realized treatments up to and including decision time $t$ for individual $j \in [G]:=\{1,\ldots, G\}$ in the cluster.   Let $\bar a_{t} = (\bar a_{t,1}, \ldots, \bar a_{t,G})$ denote the set of realized treatments up to and including decision time $t$ for all individuals in the cluster and $\bar a_{t,-j} = \bar a_t \backslash \bar a_{t,j}$ to denote this set with the $j$th individual's realized treatments removed. $Y_{t,\Delta,j} (\bar a_{t+\Delta-1}) \in \{0,1\}$ denotes the potential response of individual $j \in [G]$ under a treatment sequence $\bar a_{t+\Delta-1}$, which may depend on the realized treatments for all individuals in the cluster.

\subsection{Direct causal excursion effects}
At the cluster level, our interest lies in the effect of providing treatment versus not at time $t$ on a random individual $j$ in the cluster $[G]$.  The corresponding solutions should be constructed by comparing the potential outcomes $Y_{t,\Delta,j} (\bar a_{t,\Delta,-j}, (\bar a_{t-1,j}, a_{t,j}=1, a_{t+1:(t+\Delta-1),j}))$ and $Y_{t,\Delta,j} (\bar a_{t,\Delta,-j}, (\bar a_{t-1,j}, a_{t,j}=0, a_{t+1:(t+\Delta-1),j}))$. The ``fundamental problem of causal inference''~\citep{Rubin, pearl2009} is that these potential outcomes on the same individual cannot be observed simultaneously, so to define treatment effects, we take averages of potential relative risks. 

Let $S_t (\bar a_{t-1})$ denote a vector of moderator variables chosen from $H_t (\bar a_{t-1})$, the cluster-level history up to decision point $t$.  We write $S_t (\bar A_{t-1}) = \left( S_{t,j} (\bar A_{t-1}), S_{t,-j} (\bar A_{t-1}) \right)$ to clarify that the potential moderator variables can contain both information on the selected individual as well as other individuals in the cluster. Then the \emph{direct causal excursion effect}, denoted by $\beta_{\mathbf{p},\pi,\Delta} (t;s)$, can be defined as
\begin{equation}
\label{eq:directrelrisk}
\beta_{\mathbf{p},\pi,\Delta} (t;s)  = \log  \frac{\E_{\mathbf{p},\pi} \left[ Y_{t,\Delta,J} (\bar A_{t+\Delta-1,-J}, (\bar A_{t-1,J}, 1,\tilde{A}_{t+1:(t+\Delta-1),J})) | S_t (\bar A_{t-1})=s\right]}{\E_{\mathbf{p},\pi} \left[ Y_{t,\Delta,J} (\bar A_{t+\Delta-1,-J}, (\bar A_{t-1,J}, 0,\tilde{A}_{t+1:(t+\Delta-1),J})) | S_t (\bar A_{t-1})=s\right]},
\end{equation}
where $J$ is a uniformly distributed random index defined on $[G]$. The expectation is over the potential outcomes $Y_{t,\Delta,J}(\cdot)$, the set of randomized treatments~($\bar A_{t+\Delta-1,-J} \sim \mathbf{p}$, $\bar A_{t-1,J} \sim \mathbf{p}$, $\tilde{A}_{t+1:(t+\Delta-1),J} \sim \pi$), and the random index $J$. By ``direct'', we mean that we are only studying the relationship between the treatment realization $a_{t,j}$ and outcome $Y_{t,\Delta,j}$, with all previous and future treatments on the individual and all treatments on other members of the cluster are fixed to a specific value. Different choices of variables in $S_t (\bar A_{t-1})$ address a variety of scientific questions. A primary analysis, for instance, may focus on the fully marginal proximal effects and therefore set $S_t (\bar A_{t-1}) = \emptyset$. A second analysis may focus on assessing the effect conditional on variables only related to the individual indexed by $J$ and set $S_t (\bar A_{t-1}) = X_{t,J} (\bar A_{t-1,J})$.  A third analysis may consider cluster-level effect moderators, for example $S_t (\bar A_{t-1}) = G^{-1} \sum_j X_{t,j} (\bar A_{t-1,j})$.

\begin{remark}
Treatment micro-randomization could also take place at the cluster level. For example, as an alternative to focusing on the individual-level direct treatment effect as in \eqref{eq:directrelrisk}, one could contrast providing treatment to every member in cluster $[G]$ at time $t$ with not providing treatment to anyone, which represents a cluster-level direct effect. Domain scientists can use this setup to investigate whether community treatment allocation affects the aggregated outcome of the community as a whole. 
\end{remark}

\begin{remark}

 
Prior work by \citet{shi2021} constructed a framework where causal excursion effects are considered under potential cluster-level treatment effect heterogeneity and interference. The causal estimands target a linear contrast because the outcome of interest is continuous. As regards binary outcomes, however, relative risk is often more scientifically meaningful.
\end{remark}

\begin{remark}
The advantage of relative risk over odds ratio is that it is collapsible, i.e., there exist weights for arbitrarily specified moderators that yield a weighted average of the moderated effects \citep{pearl2009}. As such, in the context of time-varying moderated treatment effects, defining causal excursion effects as relative risks on a log scale is natural.
\end{remark}

\subsection{Pairwise indirect causal excursion effects}

Of secondary interest is a pairwise indirect effect. This is meant to measure the indirect effect on individual $j$ (who is not given a treatment) while a random treatment allocation on patient $j^\prime$ is implemented.  This pairwise within-cluster treatment interference can be assessed by comparing $Y_{t,\Delta,j} (\bar a_{t+\Delta-1,-\{j, j^\prime\}}, (\bar a_{t-1,j}, 0,a_{t+1:(t+\Delta-1),j}), (\bar a_{t-1,j^\prime}, 1,a_{t+1:(t+\Delta-1),j^\prime}))$ and $Y_{t,\Delta,j} (\bar a_{t+\Delta-1,-\{j, j^\prime\}}, (\bar a_{t-1,j}, 0,a_{t+1:(t+\Delta-1),j}), (\bar a_{t-1,j^\prime}, 0,a_{t+1:(t+\Delta-1),j^\prime}))$. 


As these two outcomes cannot be observed simultaneously, we consider averages of potential risks in the definition. The \emph{pairwise indirect causal excursion effect}, denoted by $\beta^{(IE)}_{\mathbf{p},\pi,\Delta} (t;s) $, is defined as:
\begin{equation}
\label{eq:indirecteffect}
\log  \frac{\mathbb{E}_{\mathbf{p},\pi} \left[ Y_{t,\Delta,J} (\bar A_{t+\Delta-1,-\{J, J^\prime\}}, (\bar A_{t-1,J}, 0,\tilde A_{t+1:(t+\Delta-1),J}), (\bar A_{t-1,J^\prime}, 1,\tilde A_{t+1:(t+\Delta-1),J^\prime})) | S_t (\bar A_{t-1})=s\right]}{\mathbb{E}_{\mathbf{p},\pi} \left[ Y_{t,\Delta,J} (\bar A_{t+\Delta-1,-\{J, J^\prime\}}, (\bar A_{t-1,J}, 0,\tilde A_{t+1:(t+\Delta-1),J}), (\bar A_{t-1,J^\prime}, 0,\tilde A_{t+1:(t+\Delta-1),J^\prime})) | S_t (\bar A_{t-1})=s\right]},
\end{equation}
where $J^\prime$ is uniformly distributed random index on the set $[G] \backslash \{J\}$. The expectation is over the potential outcomes of $Y_{t,\Delta,J}(\cdot)$, the set of randomized treatments~($\bar A_{t+\Delta-1,-\{J,J^\prime\}}\sim \mathbf{p}$, $\bar A_{t-1,J}\sim \mathbf{p}$, $\bar A_{t-1,J^\prime}\sim \mathbf{p}$, $\tilde A_{t+1:(t+\Delta-1),J} \sim \pi$, and $\tilde A_{t+1:(t+\Delta-1),J^\prime} \sim \pi$), and the random indices $J$ and $J^\prime$. The moderators can be written as $S_t (\bar A_{t-1}) = ( S_{t,J} (\bar A_{t-1}), S_{t,J^\prime} (\bar A_{t-1}),$ $ S_{t,-\{J, J^\prime \}} (\bar A_{t-1}))$ to clarify that the variables can contain both information on the two selected individuals as well as others in the cluster. Note that a similar effect can be defined when the individual $j$ does receive treatment. For now, we focus on the effect defined by~\eqref{eq:indirecteffect}.

\begin{remark} \normalfont
 The effect defined by \eqref{eq:indirecteffect} extends the pairwise indirect causal excursion effects in \cite{shi2021} to adjust for binary outcomes, and generalizes the \textit{group average indirect causal effect} from \cite{tchetgen2012causal} to a \textit{cluster-level causal excursion effect} that allows for moderation and time-varying treatments. The pairwise contrast is over two individuals within the same cluster, where one is fixed to receive no treatment, and vary the other individual's treatment allocation. It is important to point out that this particular definition is only one among a variety of possible definitions of the ``indirect effect'', which is of scientific interest and could be reasonably estimated. We could theoretically extend the setup beyond pairwise indirect effect, but this would be less relevant in our scientific context, and the power to detect such effects would also be low.
\end{remark}

\subsection{Identification}

To express the causal excursion effect in terms of the observed data, we make the following assumptions \citep{robins1986new}:

\begin{assumption} \normalfont
  \label{consistency}
  We assume consistency, positivity, and sequential ignorability:
  \begin{itemize}
  \item Consistency: For each~$t \leq T$ and $j \in [G]$,
    $\{Y_{t,\Delta,j} (\bar{A}_{t+\Delta-1} ), O_{t,j} (\bar A_{t-1}), A_{t,j} (\bar{A}_{t-1} )\}  = \{Y_{t, \Delta, j}, O_{t,j}, A_{t,j}\}$, where 
    $\bar A_{t}$ is \emph{joint} treatment history across all individuals in the cluster,
    i.e., observed values equal the corresponding potential outcomes;
  \item Positivity: if the joint density~$\{ H_t = h, A_t = a\}$ is greater
    than zero, then~$\mathbb{P} (A_t = a_t \given H_t = h_t ) > 0$.
  \item Sequential ignorability: for each~$t \leq T$ and $j \in [G]$, the
    potential outcomes,\\ $\{Y_{t,1,j} ( \bar a_{t}), O_{t+1,j}(\bar a_{t}),A_{t+1,j}( \bar a_{t}), \ldots,
    Y_{T,1, j} (\bar a_{T})\}_{j \in [G], \bar a_{T}\in \{0,1\}^{T\times G}}$, are independent of~$A_{t,j}$ conditional on the observed history~$H_t$.
  \end{itemize}
\end{assumption}

In an MRT, because the treatment is sequentially randomized with known probabilities bounded away from 0 and 1, positivity and sequential ignorability are satisfied by design. Here, we allow that individual treatment randomization probabilities depend on cluster-level history. Consistency is a necessary assumption for linking the potential outcomes as defined here to the data. Since an individual's outcomes may be influenced by the treatments provided to other individuals in the same cluster, consistency holds due to our use of a cluster-based conceptualization of potential outcomes as seen in \cite{hong2006evaluating} and \cite{vanderweele2013mediation}

\begin{lemma}
  \label{lemma:cond_effect}
  Under assumption~\ref{consistency}, the moderated direct treatment effect~$\beta_{\mathbf{p},\pi,\Delta} (t;s)$ is equal to
  $$
  \log  \frac{\mathbb{E} \left[  \mathbb{E} \left[W_{t,\Delta,J} Y_{t,\Delta,J} | H_{t}, A_{t,J} = 1 \right] | S_{t,J} =s\right]}{\mathbb{E} \left[  \mathbb{E} \left[W_{t,\Delta,J} Y_{t,\Delta,J} | H_{t}, A_{t,J} = 0 \right] | S_{t,J}=s \right]}, 
  $$
  where each expectation is with respect to the distribution of the data collected using the treatment assignment probabilities.

  Under assumption~\ref{consistency}, the moderated indirect treatment effect~$\beta^{(IE)}_{\mathbf{p},\pi,\Delta} (t;s)$ is equal to
  $$
  \log  \frac{\mathbb{E} \left[ \mathbb{E} \left[W_{t,\Delta,J,J^\prime} Y_{t,\Delta,J} | H_{t}, A_{t,J} = 0, A_{t,J^\prime} = 1 \right] | S_{t,J,J^\prime} =s \right]}{\mathbb{E} \left[ \mathbb{E} \left[W_{t,\Delta,J,J^\prime} Y_{t,\Delta,J} | H_{t}, A_{t,J} = 0, A_{t,J^\prime} = 0 \right] | S_{t,J,J^\prime} =s\right]}, 
  $$
\end{lemma}
\noindent where $W_{t,\Delta,J,J^\prime} = \prod_{u=t+1}^{t+\Delta-1} \pi_u(A_{u,j},A_{u,j^\prime}\given H_u)/p_u(A_{u,j},A_{u,j^\prime}\given H_u)$.  The moderators $S_{t,J}$ and $S_{t,J,J^\prime}$ can contain information on the selected individuals as well as others in the same cluster. Proof of Lemma~\ref{lemma:cond_effect} can be found in the Appendix~\ref{app:techdetails}.

\section{Estimation}
\label{section:estimation}

Motivated by the identification result in Lemma~\ref{lemma:cond_effect}, we consider estimation of direct and pairwise indirect effects using clustered MRT data.

\subsection{Direct Effect Estimation}
\label{sec:directeffect}

We make the following assumptions regarding the direct treatment effect specification:
\begin{assumption} \normalfont
\label{ass:directeffect}
Suppose that the direct causal excursion effect $\beta_{\mathbf{p},\pi,\Delta} (t;s) = f_t(S_t)^\top \beta^\star$ for some $q$-dimensional parameter $\beta^\star$, and $f_t(S_t) \in \mathcal{R}^q$ is a feature vector depending only on $S_t$ at decision point $t$.
\end{assumption}
This model allows for time-varying effects. The estimation method described below readily generalizes to situations where the parametric model has a known functional form that may be nonlinear; the use of a linear model here enhances presentation clarity. Under Assumption~\ref{ass:directeffect}, a consistent estimator $\hat{\beta}$ can be obtained by solving the following estimating equations:
\begin{align}
\label{eq:directloglin}
    \mathbb{P}_M \Bigg[ \frac{1}{G_m} \sum_{j=1}^{G_m} \sum_{t=1}^{T-\Delta +1}  W_{t,j} & W_{t,\Delta,j}e^{- A_{t,j} f_t(S_{t,j})^\top \beta} \left( Y_{t,\Delta,j} - e^{g_t(H_t)^\top \alpha + A_{t,j} f_t (S_{t,j})^\top \beta} \right) \nonumber \\ &\times \left( \begin{array}{c} g_t(H_t) \\ (A_{t,j} - \tilde p_t (A_{t,j}=1 | S_{t,j}) ) f_t(S_{t,j}) \end{array} \right) \Bigg]=0,
\end{align}
where $M$ is the total number of clusters, and $G_m$ is the size of cluster $m$. $\mathbb{P}_M$ is defined as the average of a function over the sample, which in this context is the sample of \emph{clusters}. $\exp\{g_t(H_t)^\top\alpha\}$ is a working model for $\E\{Y_{t,\Delta,J}(\Bar A_{t-1,J},0,\tilde A_{t+1:t+\Delta-1,J})|H_{t},A_{t,J}=0\}$. The weighting and centering, together with the factor $\exp\{-A_t f_t(S_t)^\top \beta\}$, makes the resulting estimator for $\beta$ consistent even when the working model $\exp\{g_t(H_t)^\top\alpha\}$ is misspecified. This new estimation method is termed as ``cluster-based estimator of the marginal excursion effect'' (C-EMEE). In Appendix~\ref{app:asymptotics}, we prove the following result:
\begin{lemma}
\label{lemma:asymnorm}
Under Assumption~\ref{ass:directeffect}, then, given invertibility and moment conditions, the estimator $\hat \beta$ that solves \eqref{eq:directloglin} satisfies $\sqrt{M} \left(\hat \beta - \beta^\star \right) \to N(0, Q^{-1} W Q^{-1})$, and 
$$
Q = \E \left[\sum_{t=1}^{T-\Delta +1}   e^{g_t(H_t)^\top \alpha^\star} \tilde{p}_t(1|S_t)(1-\tilde{p}_t(1|S_t))f_t(S_t)f_t(S_t)^\top   \right],
$$
\begin{align*}
    W =  \mathbb{E} \Bigg[ \sum_{t=1}^{T-\Delta +1}  W_{t,\Delta,J} W_{t,J}  & \tilde\epsilon_{t,\Delta,J} ( A_{t,J} - \tilde p_t( 1 | S_{t,J} )) f_t (S_{t,J}) \times \\ &\sum_{t=1}^T W_{t,\Delta,J^\prime} W_{t,J^\prime} \, \tilde\epsilon_{t,\Delta,J^\prime} ( A_{t,J^\prime} - \tilde p_t( 1 | S_{t,J^\prime} )) f_t (S_{t,J^\prime})^\top  \Bigg],
\end{align*}
where $\tilde\epsilon_{t,\Delta,j} = e^{-A_{t,j} f_t(S_{t,j})^\top \beta^\star} \times \left(Y_{t,\Delta,j} - e^{g_t(H_t)^\top \alpha^\star + A_{t,j} f_t (S_{t,j})^\top \beta^\star}\right)$, and $\alpha^\star$ solves Equation~(\ref{eq:directloglin}). Both $J$ and $J^\prime$ are randomly sampled indices from the same cluster.
\end{lemma}

\noindent In practice, plug-in estimates $\hat Q$ and $\hat W$ are used to estimate the covariance structure. Appendix~\ref{app:ssa} presents small sample size adjustments that are used in the analysis.

\subsection{Connection to the Standard MRT Analysis}
\label{section:samesies}

In standard MRTs, the individual is the unit of interest. Here, the cluster is the unit of interest \citep{shi2021}.  It is clear from Lemma~\ref{lemma:asymnorm} that if the cluster-size is one then we recover the estimators and asymptotic theory for EMEE. In addition, we explore if there are other cases in which two methods yield equivalent inference results, thus allowing us to develop a general connection between them. 

Lemma~\ref{lemma:samesies} below implies that if cluster-level variations in proximal outcomes are independent of treatment option, then both methods are equivalent. In contrast, if these types of cluster-level variation differ by treatment status, then C-EMEE and EMEE will no longer be equivalent. Proof of Lemma~\ref{lemma:samesies} can be found in Appendix~\ref{app:samesies}. 

\begin{lemma}
\label{lemma:samesies}
Consider the direct effect when the moderator is defined on the individual (i.e., $S_{t,j}$), and the randomization probabilities only depend on the individual's observed history, (i.e., $p(A_{t,j} | H_t) = p(A_{t,j} | H_{t,j})$).  If the cluster size is constant across clusters (i.e., $G_m \equiv G$), then the point estimates from \eqref{eq:mrtstandard} and~\eqref{eq:directloglin} are equal for any sample size. Moreover, if 
\begin{equation}
    \label{eq:samesiescondition}
    \E \left[ \E \left[ W_{t,\Delta,j}\tilde\epsilon_{t,\Delta,j} \times W_{t^\prime,\Delta, j^\prime} \tilde\epsilon_{t^\prime,\Delta, j^\prime} \given H_{t,j}, A_{t,j}=a, H_{t^\prime,j^\prime}, A_{t^\prime,j^\prime} = a^\prime \right] | S_{t,j}, S_{t^\prime,j^\prime} \right] = \psi(S_{t,j}, S_{t^\prime,j^\prime}),
\end{equation}
for some function $\psi$, i.e., the marginal residual correlation doesn't depend on $a$ and $a^\prime$, then the estimators share the same asymptotic variance.
\end{lemma}

To illustrate Lemma \ref{lemma:samesies}, consider $\Delta=1$ and for participant $j$ at decision time~$t$, suppose the generative model for the proximal outcome is:
$$
\E[Y_{t,1,j}] =\mathbf{c}(S_{t,j}) \exp\left\{g_t(H_t)^\top \alpha + \underbrace{X_{t,j}^\top e_g}_{(I)} + A_{t,j} \left(f_t (H_t)^\top \beta+ \underbrace{X_{t,j}^\top b_g}_{(II)}\right)  \right\},
$$
where $\mathbf{c}(S_{t,j})$ does not contain any treatment terms.
$(I)$ and $(II)$ are random effects with design matrix $X_{t,j}$. Random effects in $(I)$ allow for cluster-level variation in proximal outcomes regardless of the treatment allocation, while random effects in $(II)$ allow for cluster-level variation in the fully-conditional treatment effect. Given the above generative model, sufficient condition \eqref{eq:samesiescondition} holds if $b_g \equiv 0$, i.e., when the treatment effect does not exhibit cluster-level variation. 

For this reason, we refer to \eqref{eq:samesiescondition} as a \emph{treatment-effect heterogeneity condition} for binary outcomes. When clusters exhibit treatment-effect heterogeneity, the proposed approach is necessary for assessing direct effects rather than standard MRT analyses. In Section~\ref{section:sims}, we conduct simulations to support this statement. 

\subsection{Pairwise Indirect Causal Excursion Effect Estimation}
We make the following assumptions regarding the pairwise indirect treatment effect specification:
\begin{assumption}
\label{ass:indirecteffect}
Assume the indirect treatment effect of interest $\beta^{(IE)}_{\mathbf{p},\pi,\Delta} (t;s) = f_t(S_t)^\top \beta^{\star \star}$ for some $q$-dimensional parameter $\beta^{\star \star}$, and $f_t(S_t) \in \mathcal{R}^q$ is a feature vector depending only on the state at decision point $t$, $S_t$.

\end{assumption}
We now consider the inference on the unknown $q$-dimensional parameter $\beta^{\star \star}$. To do so, we define a new weight as $W_{t,j,j^\prime} = \frac{\tilde p(A_{t,j},A_{t,j^\prime}|S_t)}{p_t(A_{t,j},A_{t,j^\prime}|H_t)}$, where $\tilde p_t(a,a^\prime\given S_t) \in (0,1)$ is arbitrary as long as it does not depend on terms in $H_t$ other than $S_t$, and $p_t(A_{t,j},A_{t,j^\prime} \given H_t)$ is the marginal probability that individuals $j$ and $j^\prime$ receive treatments $A_{t,j}$ and $A_{t,j^\prime}$ respectively given $H_t$. Here we consider an estimator~$\hat{\beta}^{(IE)}$ of $\beta^{\star \star}$ which solves the following set of estimating equations:
\begin{align}
\label{eq:indirectloglin}
& \mathbb{P}_M \Bigg[ \frac{1}{G_m(G_m-1)} \sum_{j\neq j'}^{G_m} \sum_{t=1}^{T-\Delta +1}  W_{t,j,j^\prime}W_{t,\Delta,j,j^\prime} e^{- (1-A_{t,j})A_{t,j^\prime} f_t(S_{t,j,j^\prime})^\top \beta}~~~~ \times \\ & \left( Y_{t,\Delta,j} - e^{g_t(H_t)^\top \alpha + (1-A_{t,j})A_{t,j^\prime} f_t (S_{t,j,j^\prime})^\top \beta} \right)\nonumber  \left( \begin{array}{c} g_t(H_t) \\ (1-A_{t,j})(A_{t,j^\prime} - \tilde p_t^\star (1 | S_{t,j,j^\prime}) ) f_t(S_{t,j,j^\prime})\end{array} \right) \Bigg] =0, 
\end{align}
where $\tilde p_t^\star (1| S_{t,j,j^\prime}) = \frac{\tilde p_t (0,1 | S_{t,j,j^\prime}) }{\tilde p_t (0,0 | S_{t,j,j^\prime})+\tilde p_t (0,1 | S_{t,j,j^\prime})}$. If an individual's randomization probabilities only depends on their own observed history, then $\tilde p_t^\star (1 | S_{t,j,j^\prime}) = \tilde p_t (A_{t,j^\prime} = 1 | S_{t,j^\prime})$, and the weight can be simplified to $W_{t,j,j^\prime} = W_{t,j}\times W_{t,j^\prime}$ and  $W_{t,\Delta,j,j^\prime} = W_{t,\Delta,j}\times W_{t,\Delta,j^\prime}$. In Appendix \ref{app:asymptotics}, we proved the following result:

\begin{lemma}
\label{lemma:indasymnorm}
Under assumption~\ref{ass:indirecteffect}, invertibility and moment conditions, the estimator $\hat{\beta}^{(IE)}$ that solves \eqref{eq:indirectloglin} satisfies $\sqrt{M} \left(\hat{\beta}^{(IE)} - \beta^{\star \star} \right) \to N(0, Q^{-1} W Q^{-1})$, where
$$
Q =  \E \left[\sum_{t=1}^{T-\Delta +1}  e^{g_t(H_t)^\top \alpha^{\star \star}} \big( \tilde p_t (0,1 | S_t)+\tilde p_t (0,0 | S_t)\big) \tilde p^\star_t (1 | S_t)(1-\tilde p^\star_t (1 | S_t))    f_t(S_t) f_t(S_t)^\top \right],
$$
\begin{align*}
W =  &\mathbb{E} \bigg[ \sum_{t=1}^{T-\Delta +1}  W_{t,J,J^\prime} W_{t,\Delta, J,J^\prime} \tilde\epsilon_{t,J,J^\prime} (1-A_{t,J})( A_{t,J^\prime} - \tilde p_t^\star( 1 | S_{t,J,J' } )) f_t (S_{t,J,J' }) \\
&\times \sum_{t=1}^T W_{t,\tilde J,\tilde J^\prime} \times W_{t, \Delta, \tilde J, \tilde J^\prime} \tilde\epsilon_{t, \tilde J, \tilde J^\prime} (1-A_{t,\tilde J}) ( A_{t,\tilde J^\prime} - \tilde p_t^\star( 1 | S_{t,\tilde J, \tilde J' } )) f_t (S_{t,\tilde J, \tilde J' })^\top  \bigg].
\end{align*}
 where $\tilde \epsilon_{t,j,j^\prime}= e^{-A_{t,j} f_t(S_{t,j,j^\prime})^\top \beta^{\star \star}} \Big(Y_{t,\Delta,j} - e^{g_t(H_t)^\top \alpha^{\star \star} + (1-A_{t,j})A_{t,j^\prime} f_t (S_{t,j,j^\prime})^\top \beta^{\star \star}}\Big)$, and $\alpha^{\star \star}$ solves Equation~(\ref{eq:indirectloglin}). Both $(J,J^\prime)$ and $(\tilde J, \tilde J^\prime)$ are independently, randomly sampled pairs from the same cluster.

\end{lemma}


\section{Simulations}
\label{section:sims}


The simulation design is a generalization of the simulation experiments in \cite{qian2020estimating}. Our simulation study demonstrates the performance of the proposed methods by generating data with predetermined cluster structures. The purpose of this study is to demonstrate that when within-cluster interference and cluster-level treatment effect heterogeneity (discussed in Section \ref{section:samesies}) exist, the proposed C-EMEE method produces a consistent estimator with nominal coverage probability, whereas the EMEE may not. 
\subsection{Simulation setup}

To evaluate the performance of the proposed inferential procedure, we use the following generative model. The states $Z_{t,j}$ are generated using a first-order Markov chain within each subject. The treatment randomization probability is constant with $p(A_t=1|H_t)=0.2$. We focus on lag-1 proximal response ($\Delta=1$), and the outcome $Y_{t,1,j}$ is generated from a Bernoulli distribution with
\begin{equation}
\label{eq:generativemodel}
    \E (Y_{t,1,j}\given H_t,A_{t,j}) = \left\{0.1 \mathbf{1}_{Z_{t,j}=0}+0.25 \mathbf{1}_{Z_{t,j}=1}+0.2 \mathbf{1}_{Z_{t,j}=2}  \right\}e^{A_{t,j}(0.1+0.3Z_{t,j})}.
\end{equation}

Here, $Z_{t,j}$ moderates the conditional treatment effect and the true conditional treatment effect is:
$
\beta_{\mathbf{p},1}(t;H_t) = 0.1+0.3Z_{t,j}.
$ Additional details about the simulation setup can be found in Appendix \ref{app:markovchain}. 


\subsection{Simulation Scenarios}

In the following, we consider four simulation scenarios that extend the generative model described above. All four scenarios concern the estimation of the marginal proximal treatment effect $\beta_{0}^\star$. In each case, we report the bias, standard errors (SE), root mean squared error (RMSE) of $\hat{\beta}_{0}$, and 95\% confidence interval coverage probabilities (CP) across 1000 replicates. We adopt the small sample correction technique in \cite{mancl2001covariance} to modify the variance estimator. Also, as in \cite{liao2016sample}, we use critical values from a $t$ distribution. In particular, for a known $p$-dimensional vector $c$, to test the null hypothesis $c^\top \beta = 0$ or to form two-sided confidence intervals, we use the critical value $t^{-1}_{n-p-q}(1-\xi/2)$ where $p,q$ are the dimensions of $\alpha,\beta$; respectively, and $\xi$ is the significance level. We vary the number of clusters and cluster size and compare the proposed cluster-based method C-EMEE to EMEE. As we are using clusters of equal size, we expect both methods to produce the same point estimation. Two methods will give different estimations if cluster sizes are different.  

\noindent {\bf Simulation Scenario I}. The generative model is as follows:
\begin{equation}
    \E (Y_{t,1,j}\given H_t,A_{t,j}) = \left\{0.1 \mathbf{1}_{Z_{t,j}=0}+0.25 \mathbf{1}_{Z_{t,j}=1}+0.2 \mathbf{1}_{Z_{t,j}=2}  \right\}e^{A_{t,j}(0.1+0.3Z_{t,j})+e^\prime_g}.
\end{equation}

In particular, the present scenario concerns the estimation of $\beta_{0}^\star$ when an important individual-level moderator exists, and the proximal outcomes for subjects in the same cluster share a random cluster-level intercept term that does not interact with treatment. To prevent this intercept from shifting the expected outcome, we modified it by two steps: First, we draw $e_g$ from a truncated normal distribution with $\mu =0,\sigma^2=0.5, a = -1, b=1$. Second, we shift $e_g$ to $e_g^\prime$ to ensure that $\E[e^{e^\prime_g}]=\E[e^\mu]= 1$, in which case, $e^\prime_g = e_g - \frac{\sigma^2}{2}-\log \frac{\Phi(\frac{b}{\sigma}-\sigma)-\Phi(\frac{a}{\sigma}-\sigma)}{\Phi(\frac{b}{\sigma})-\Phi(\frac{a}{\sigma})}$.

The resulting marginal excursion effect is:
\begin{equation}
\label{eq:beta_0star}
    \beta^\star_0 = \log  \frac{\E\left\{\E(Y_{t,1,j}\given H_t,A_{t,j}=1)\right\}}{\E\left\{\E(Y_{t,1,j}\given H_t,A_{t,j}=0)\right\}}=0.477.
\end{equation} 

Table~\ref{tab:simresults} shows that both EMEE and the proposed C-EMEE approach achieve nearly unbiasedness and proper coverage. This is in line with our theoretical results in Section~\ref{section:samesies} stating asymptotic equivalence of the two procedures under no cluster-level treatment heterogeneity. This demonstrates that the performance of the EMEE approach is not impacted by cluster-level correlation that does not interact with treatment.

\noindent {\bf Simulation Scenario II}. In the second scenario, we extend the above generative model to include a random cluster-level intercept term that interacts with treatment. Similar to $e_g$ and $e_g^\prime$, $b_g$ follows a truncated normal distribution with $\mu=0, \sigma=0.5, a=-1, b=1$, and $b^\prime_g = b_g- \frac{\sigma^2}{2}-\log \frac{\Phi(\frac{b}{\sigma}-\sigma)-\Phi(\frac{a}{\sigma}-\sigma)}{\Phi(\frac{b}{\sigma})-\Phi(\frac{a}{\sigma})}$ is a random-intercept term within the treatment effect for cluster $g$. The generative model then takes the form of:
\begin{equation}
    \E (Y_{t,1,j}\given H_t,A_{t,j}) = \left\{0.1 \mathbf{1}_{Z_{t,j}=0}+0.25 \mathbf{1}_{Z_{t,j}=1}+0.2 \mathbf{1}_{Z_{t,j}=2}  \right\}e^{A_{t,j}(0.1+0.3Z_{t,j}+b^\prime_g)}.
\end{equation}

\begin{figure}
 \centerline{\includegraphics[width=4in]{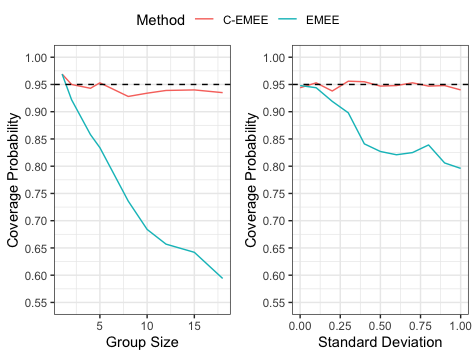}}
 \caption{{\bf Left:} Cluster size  ($G$) versus nominal coverage probability with fixed total sample size. {\bf Right:} Standard deviation of random intercept $b_g^\prime$ versus nominal coverage probability.  
    }
    \label{fig:undercoverage}
\end{figure}

The fully marginal excursion effect is equal to \eqref{eq:beta_0star}.
In Table~\ref{tab:simresults} we observe that if cluster-level random effects interact with treatment  (i.e. $b^\prime_g \not \equiv 0$), then both methods produce nearly unbiased estimates of $\beta_{0}$ but only the proposed method achieves the nominal 95\% coverage probability. To further demonstrate this, Figure~\ref{fig:undercoverage} presents nominal coverage as a function of the standard deviation of the random effect $b^\prime_g$ and cluster size respectively.  This shows that the coverage probability of the EMEE method decays rapidly while the proposed method achieves the nominal 95\% coverage probability for all choices of the standard deviation of $b^\prime_g$ and cluster sizes. Note that even when cluster size is $5$ (i.e., small clusters), the nominal coverage drops below 85\%.

\noindent {\bf Simulation Scenario III}.
In the third scenario, we assume the treatment effect for an individual depends on the average state of all individuals in the cluster. We simulate data under a ground truth with cluster-level moderator $\bar Z_{t,g}= \frac{1}{G_g}\sum_{j=1}^{G_g}Z_{t,j}$ for cluster $g$:
\begin{equation}
    \E (Y_{t,1,j}\given H_t,A_{t,j}) = \left\{0.1 \mathbf{1}_{Z_{t,j}=0}+0.25 \mathbf{1}_{Z_{t,j}=1}+0.2 \mathbf{1}_{Z_{t,j}=2}  \right\}e^{A_{t,j}(0.1+0.3 \Bar{Z}_{t,g}+b^\prime_g)}.
\end{equation}

The true marginal excursion effect here is:
\begin{equation}
    \beta^\star_0 = \log  \frac{\E\left\{\E(Y_{t,1,j}\given H_t,A_{t,j}=1)\right\}}{\E\left\{\E(Y_{t,1,j}\given H_t,A_{t,j}=0)\right\}}=0.4.
\end{equation}

The proposed estimator again achieves the nominal 95\% coverage probability while the EMEE method does not (see Scenario III, Table~\ref{tab:simresults}).
\vspace{-0.5cm}

\begin{table}
\centering
\caption{ C-EMEE and EMEE comparison over Scenarios I: ; II:; and III;}
\label{tab:simresults}
\begin{tabular}{cccccccc}
\hline
Scenario & Estimator & \# of Clusters & Cluster Size & Bias & SE & RMSE & CP \\ \hline
\multirow{12}{*}{I} & C-EMEE & \multirow{2}{*}{25} & \multirow{2}{*}{5} & \multirow{2}{*}{$2.73 \times 10^{-3}$} & 0.069 & 0.069	 & 0.949 \\ 
& EMEE & & &   & 0.069 & 0.069 & 0.948 \\   
& C-EMEE & \multirow{2}{*}{25} & \multirow{2}{*}{10} & \multirow{2}{*}{$-6.40 \times 10^{-4}$} & 0.048 & 0.047 & 0.945 \\ 
& EMEE & & &  & 0.049 & 0.049 & 0.941 \\  
& C-EMEE & \multirow{2}{*}{50} & \multirow{2}{*}{10} & \multirow{2}{*}{$-3.96 \times 10^{-4}$} & 0.034 & 0.034 & 0.938 \\ 
& EMEE & & &   & 0.034 & 0.034 & 0.940 \\  
& C-EMEE & \multirow{2}{*}{50} & \multirow{2}{*}{20} & \multirow{2}{*}{$-7.86 \times 10^{-4}$} & 0.024 & 0.024 & 0.957 \\ 
& EMEE & & &  & 0.024 & 0.024 & 0.954 \\  
& C-EMEE & \multirow{2}{*}{100} & \multirow{2}{*}{20} & \multirow{2}{*}{$-1.15 \times 10^{-3}$} & 0.017 & 0.016 & 0.949 \\ 
& EMEE & & &  & 0.016 & 0.016 & 0.950 \\  
& C-EMEE & \multirow{2}{*}{100} & \multirow{2}{*}{25} & \multirow{2}{*}{$-2.58 \times 10^{-4}$} & 0.015 & 0.016  & 0.941 \\
& EMEE & & &  & 0.015 & 0.016 & 0.937 \\ \hline
\multirow{12}{*}{II} & C-EMEE & \multirow{2}{*}{25}  & \multirow{2}{*}{5}& \multirow{2}{*}{$-1.09 \times 10^{-2}$} & 0.113 & 0.114 & 0.937 \\ 
& EMEE & & &   & 0.078 & 0.114 & 0.816 \\   
& C-EMEE & \multirow{2}{*}{25} & \multirow{2}{*}{10} & \multirow{2}{*}{$-7.65 \times 10^{-3}$} & 0.102 & 0.101 & 	0.934 \\ 
& EMEE & & &   & 0.055 & 0.101 & 0.719 \\  
& C-EMEE & \multirow{2}{*}{50} & \multirow{2}{*}{10} & \multirow{2}{*}{$-3.51 \times 10^{-3}$} & 0.072 & 0.072 & 0.957 \\ 
& EMEE & & &  & 0.039 & 0.072 & 0.717 \\  
& C-EMEE & \multirow{2}{*}{50} & \multirow{2}{*}{20} & \multirow{2}{*}{$-2.88 \times 10^{-3}$} & 0.068 & 0.069 & 0.934 \\ 
& EMEE & & &   & 0.027 & 0.069 & 0.563 \\  
& C-EMEE & \multirow{2}{*}{100} & \multirow{2}{*}{20} & \multirow{2}{*}{$-1.83 \times 10^{-3}$}  & 0.048 & 0.048 & 0.941 \\ 
& EMEE & & &   & 0.019 & 0.048 & 0.567 \\  
& C-EMEE & \multirow{2}{*}{100} & \multirow{2}{*}{25} & \multirow{2}{*}{$-1.03 \times 10^{-3}$}  & 0.047 & 0.048 & 0.943 \\ 
& EMEE & & &  & 0.017 & 0.048 & 0.526 \\ \hline
\multirow{12}{*}{III} & C-EMEE & \multirow{2}{*}{25} & \multirow{2}{*}{5} & \multirow{2}{*}{$8.90 \times 10^{-3}$} & 0.115 & 0.111 & 0.957 \\
& EMEE & & &  & 0.080 & 0.111 & 0.840 \\   
& C-EMEE & \multirow{2}{*}{25} & \multirow{2}{*}{10} & \multirow{2}{*}{$6.29 \times 10^{-3}$}  & 0.104 & 0.104 & 0.945 \\ 
& EMEE & & &   & 0.057 & 0.104 & 0.712 \\  
& C-EMEE & \multirow{2}{*}{50} & \multirow{2}{*}{10} & \multirow{2}{*}{$8.30 \times 10^{-3}$} & 0.072 & 0.073 & 0.942 \\ 
& EMEE & & &  & 	0.040 & 0.074 & 0.728 \\  
& C-EMEE & \multirow{2}{*}{50} & \multirow{2}{*}{20} & \multirow{2}{*}{$-7.74 \times 10^{-4}$} & 0.068 & 0.067 & 0.939 \\ 
& EMEE & & &  & 0.028 & 0.067 & 0.603 \\  
& C-EMEE & \multirow{2}{*}{100} & \multirow{2}{*}{20} & \multirow{2}{*}{$3.36 \times 10^{-3}$} & 0.048 & 0.047 & 0.952 \\ 
& EMEE & & &   & 0.020 & 0.047 & 0.594 \\  
& C-EMEE & \multirow{2}{*}{100} & \multirow{2}{*}{25} & \multirow{2}{*}{$1.70 \times 10^{-3}$} & 0.047 & 0.049 & 0.943 \\ 
& EMEE & & &  & 0.018 & 0.049 & 0.524 \\ \hline
\end{tabular}
\end{table}

\vspace{-0.5cm}

\noindent {\bf Simulation Scenario IV}. The fourth scenario considers the pairwise indirect excursion effect. For individual $j$ at decision point $t$, we constructed the total effect as $TE_{t,j} =\sum_{j\neq j^\prime}A_{t,j^\prime}\beta_{20} $, where $\beta_{20}=-0.1$. The generative model is given by:
\begin{equation}
    \E (Y_{t,1,j}\given H_t,A_{t,j}) = \frac{1}{\gamma}\left\{0.1 \mathbf{1}_{Z_{t,j}=0}+0.25 \mathbf{1}_{Z_{t,j}=1}+0.2 \mathbf{1}_{Z_{t,j}=2}  \right\}e^{A_{t,j}(0.1+0.3 \Bar{Z}_{t,g}+b^\prime_g)+TE_{t,j}}.
\end{equation}

In order to prevent the newly-added total effect term from inflating the expected outcome and keep it on the same scale with previous three scenarios, we divided a regularization constant $\gamma =[pe^{\beta_{20}}+(1-p)]^{m-2}$ in front of the generative model (see Appendix \ref{app:indirecteffect_factor} for proof). This model implies a fully marginal indirect effect equal to:
\begin{equation}
    \beta^{\star\star}_0 = \log  \frac{\E\left\{\E(Y_{t,1,j}\given H_t,A_{t,j}=0,A_{t,j^\prime}=1)\right\}}{\E\left\{\E(Y_{t,1,j}\given H_t,A_{t,j}=0,A_{t,j^\prime}=1)\right\}}=-0.1.
\end{equation}
 
 Table~\ref{tab:simresults_indirect} presents the results. We see that the proposed indirect estimator exhibited nearly no bias and achieved the nominal coverage probability. 

\begin{table}[!th]
\centering
\caption{C-EMEE for estimation of indirect effects.}
\label{tab:simresults_indirect}
\begin{tabular}{ccccccc}
\hline
Scenario & \# of Clusters & Cluster Size & Bias & SE & RMSE & CP \\ \hline
\multirow{6}{*}{IV} & 25 & 5 & $-2.01\times 10^{-4}$ & 0.051 & 0.050 & 0.947 \\
& 25 & 10 & $-3.96\times 10^{-4}$ & 0.025 &  0.024 & 0.951 \\ 
& 50 & 10 & $-4.36\times 10^{-4}$ & 0.017 & 0.017 & 0.956 \\ 
& 50 & 20 &  $3.91\times 10^{-4}$ & 0.009 & 0.009 & 0.953 \\ 
& 100 & 20 & $2.89\times 10^{-4}$& 0.006 & 0.006 & 0.945 \\ 
& 100 & 25 &  $-1.36\times 10^{-4}$ & 0.005 & 0.005 & 0.951 \\ \hline
\end{tabular}
\end{table}

In addition, similar conclusion holds when $\Delta >1$, and the simulation results can be found in Appendix \ref{app:lagsimulation}. 

\section{Case Study}
\label{sec:casestudy}

In Section \ref{sec:motivatingexample}, we introduced the Intern Health Study \citep{Necamp2020}, which is a 6-month MRT exploring when to offer mHealth interventions to individuals in stressful environments to improve their mental health. According to Figure \ref{fig:clusterdif}, sending a notification can on average have a negative impact on survey completion the following day at most specialties. Our aim in this section is to reevaluate this causal effect in light of the results for all participants. We will perform inference on the lag-1 ($\Delta =1$) marginal and moderated causal excursion effects, and the data set used in the analyses contains 1562 participants.

In this paper, all analyses are conducted using daily data. Recall that an individual was randomized to receive a mobile notification with probability 0.375 every day. Because of the form of the intervention, all participants were assumed available for this intervention throughout the study; i.e., $I_{t} =1$. The binary proximal outcome $Y_{t,1}$, mood survey completion, is coded as 1 if a participant self-reported his or her mood score on that day, and 0 otherwise. The average daily mood survey completion rate is 0.344  when a notification is delivered and 0.367 when there is no notification. In the following analysis, clusters are constructed based on a subject's membership of medical specialty, because with similar working content and schedules, interns are more likely to adjust to notifications similarly. The effect of targeted notification treatment on interns' engagement with the mobile intervention study is assessed using our proposed method, C-EMEE, and compared with EMEE. 


\subsection{Direct treatment effect analysis}
In this section, we assess the marginal excursion effect as well as the effect moderation by time-varying observations. For an individual $j$, the day-in-study is coded as a subscript $t$. For all the analyses in this section, we always include the prior week's completion rate ($R_{t,j}$) and day-in-study ($t_j$) in the control variable $g_t(H_t)$, as they are prognostic of $Y_{t,1,j}$ in a preliminary generalized estimating equation. Therefore, the working model is specified as $g_t(H_t)^\top\alpha = \alpha_0 + \alpha_1 t_j + \alpha_2 R_{t,j}$.

First, we estimate the marginal treatment effect $\beta_0$ of the targeted notifications on completing self-reported mood score survey with the analysis model: 
\begin{equation*}
    \log  \frac{\E\left\{\E(Y_{t,1,j}\given H_t,A_{t,j}=1)\right\}}{\E\left\{\E(Y_{t,1,j}\given H_t,A_{t,j}=0)\right\}} = \beta_0.
\end{equation*}

Results of the fully marginal analysis are presented in Table \ref{tab:application} and compares our proposed approach against the EMEE approach from \cite{qian2020estimating}. The results show that the SE increases with increasing cluster size. Across all clustering levels explored in our study, the marginal excursion effect differs significantly from zero. This suggests that sending a notification reduces the likelihood of responding to a mood survey the next day.


Two analyses are conducted by incorporating an individual-level effect moderator $S_{t,j}$ with the model:
$$
    \log \frac{\E\left\{\E(Y_{t,1,j}\given H_t,A_{t,j}=1)\given S_t\right\}}{\E\left\{\E(Y_{t,1,j}\given H_t,A_{t,j}=0)\given S_t\right\}} = \beta_0+\beta_1 S_{t,j},
$$ 
where in the first model, $S_{t,j} = R_{t,j}$ represents a subject's completion rate during the past week, whereas in the second model, $S_{t,j} = t_j$ represents a subject's day-in-study. Estimated treatment effects moderation by prior week's completion rate or day-in-study are shown in Figure \ref{fig:two graphs} below. The shaded area represents the 95\% confidence band of the moderation effects at varying values of the moderator. In the first 50 days of the study, sending notifications had no significant impact on the daily survey completion. However, in the latter half of the study, sending a notification was significantly less likely to elicit engagement.

In light of this, it might not be ideal to overburden participants for an extended period if the notifications don't serve any therapeutic purpose. In addition, notifying highly engaged or less engaged participants will not affect their willingness to complete the mood survey. In contrast, mobile prompts negatively affect the completion status of participants with a prior week's completion rate roughly between 0.2 and 0.6. Additional analysis has been done on evaluating lag treatment effects (i.e., $\Delta > 1$), see more details in Appendix \ref{app:morecasestudy}. 

\begin{figure}
 \centerline{\includegraphics[width=5.5in]{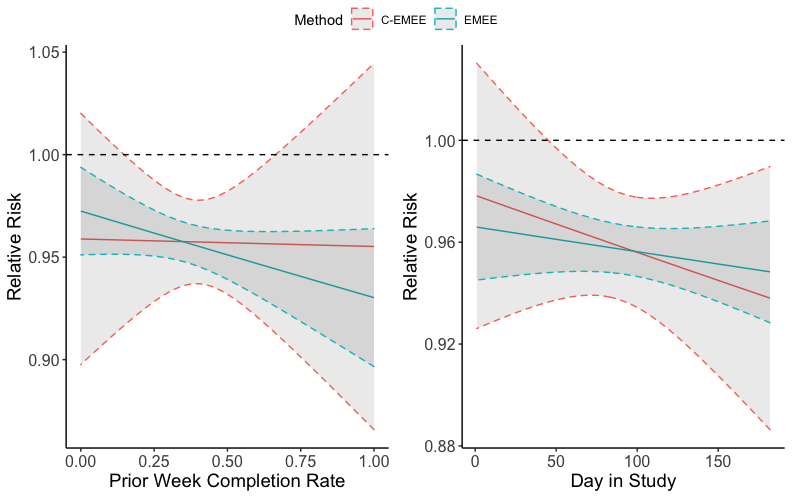}}
\caption{Moderation Analysis: \textbf{Left}: Treatment effect moderate by prior week's completion rate, and \textbf{Right}: Treatment effect moderate by day-in-study.}
\label{fig:two graphs}
\end{figure}

\subsection{Pairwise indirect treatment effect analysis}

Finally, we consider pairwise indirect effect analyses. In this analysis, clusters are constructed based on each subjects' joint membership of medical specialty and institution. This was done as interference was only likely when interns are in close geographic proximity. Here, we consider the marginal indirect effect (i.e., no moderators) when the individual did not receive the intervention at decision time $t$. Table \ref{tab:application}  presents the results. We see limited evidence of an indirect effect. 

\begin{table}[!th]
\centering
\caption{Fully marginal causal excursion effect of notifications on daily mood survey completion rate in IHS. EMEE estimates use the estimator in~\citet{qian2020estimating}, while the C-EMEE estimates use our proposed estimator.} 
\label{tab:application}
\begin{tabular}{cccccccc}
\hline
Outcome & Cluster  & Method  & Estimate & Std. Error  & p-value \\ \hline
\multirow{3}{*}{Direct Effect} & Institution $\times$ Specialty & C-EMEE  & -0.058 & 0.006  & \textbf{$<$0.05} \\ 
& Institution & C-EMEE  & -0.055 & 0.009  & \textbf{$<$0.05} \\  
& Individual & EMEE & -0.053 &  0.005  & \textbf{$<$0.05} \\ 
\hline
\multirow{1}{*}{Indirect Effect} & Institution $\times$ Specialty & C-EMEE   & -0.003  & 0.010  & 0.74\\ \hline
\end{tabular}
\end{table}


Our analysis suggests a significantly lower response rate to mood surveys when a mobile intervention is delivered compared to when it is not delivered.  We conclude that researchers should not only consider the therapeutic effect of mobile interventions but also the impact on engagement when designing mobile intervention components in future mHealth studies.


\section{Discussion}

This paper considered the statistical problem of assessing whether time-varying interventions have a causal effect on a proximal and/or lagged binary response. We defined novel direct and pairwise indirect causal excursion effects on the longitudinal binary outcome by accounting for potential within-cluster interference and between cluster treatment heterogeneity. Theoretical results, including consistency and asymptotic normality of the proposed estimators, are presented.

The methods we propose have several practical advantages over EMEE. First, our proposed approach provides more valid interval estimates, hence addresses the undercoverage issues associated with existing inferential methods, ensuring a reasonable inference for the intervention effectiveness. Second, the proposed estimands and estimation methods enable domain scientists to answer a wider variety of scientific questions. For example, it is now possible to select moderators at the cluster level, which provides deeper insight into how community aggregated characteristics affect an individual's binary outcome of interest. It also enables the estimation of peer effects, which may be of scientifical interest in, for example, epidemiology and sociology research. Finally, the proposed method is easy-to-implement, with all estimation procedures we discussed above easily implemented in R using existing statistics packages.

While this work represents a major step forward in the analysis of data from micro-randomized trials, a few directions for future research are worth considering. First we have assumed a binary treatment in this work; extension to treatment with multiple levels can refine causal understanding of categorical treatments.
Second, since there are numerous moderators that can be potentially included in $f_t(S_t)$ to improve estimation efficiency of the treatment effect, regularization methods for model selection in building the causal effect model $f_t(S_t)^\top\beta$ with high-dimensional $f_t(S_t)$ can be useful. Lastly, it is worthwhile to extend the current framework beyond clustered structures such as overlapping communities or a general network among the subjects \citep{Ogburn2014,Mealli2019}. We leave these topics for future work.



\backmatter






%

\bibliography{my_ref.bib}
\bibliographystyle{biom}









\label{lastpage}

\newpage

\section{Supporting Information}

Supplementary material includes proof of the identification result (Appendix \ref{app:techdetails}), proof of Lemma \ref{lemma:asymnorm} and Lemma \ref{lemma:indasymnorm} (Appendix \ref{app:asymptotics}), and proof of Lemma~\ref{lemma:samesies} (Appendix \ref{app:samesies}). We provide additional simulation study results in Appendix \ref{app:moresimulation} and \ref{app:lagsimulation}. An analysis on lagged outcomes in the case study is given in Appendix \ref{app:morecasestudy}.

\subsection{Technical Details}
\label{app:techdetails}

\begin{proof}[Proof of the identifiability result] It suffices to show that under Assumption \ref{consistency}, the following equation holds:
\begin{align}
    &\E \left\{Y_{ t,\Delta,J}(\Bar{A}_{t-1,J},a, \tilde{A}_{t+1:(t+\Delta-1),J})\given S_t(\Bar{A}_{t-1}), I_{t,J}(\Bar{A}_{t-1,J})=1 \right\} \nonumber \\
    = &\E \left[\E\left\{ Y_{ t,\Delta,J}\given H_t, A_{t,J}=a, I_{t,J} =1\right\}\given S_t, I_{t,J} = 1\right]
\end{align}

We have the following sequence of equality:
\begin{align*}
    &\E \left\{Y_{ t,\Delta,J}(\Bar{A}_{t-1,J},a,\tilde{A}_{t+1:(t+\Delta-1),J})\given S_t(\Bar{A}_{t-1}), I_{t,J}(\Bar{A}_{t-1,J})=1 \right\} \nonumber \\
    = &\E \left[ \E \left\{Y_{ t,\Delta,J}(\Bar{A}_{t-1,J},a,\tilde{A}_{t+1:(t+\Delta-1),J})\given H_t(\Bar{A}_{t-1}), I_{t,J}(\Bar{A}_{t-1,J})=1 \right\}\given S_t(\Bar{A}_{t-1}), I_{t,J}(\Bar{A}_{t-1,J})=1\right] \\
    = &\E \left[ \E \left\{Y_{ t,\Delta,J}(\Bar{A}_{t-1,J},a,\tilde{A}_{t+1:(t+\Delta-1),J})\given H_t  \right\}\given S_t,I_{t,J} = 1 \right]\\
    = &\E \left[ \E \left\{Y_{ t,\Delta,J}(\Bar{A}_{t-1,J},a,\tilde{A}_{t+1:(t+\Delta-1),J})\given H_t,A_{t,J}=a  \right\}\given S_t,I_{t,J} = 1\right]\\
    = &\E \left[\E\left\{ Y_{ t,\Delta,J}\given H_t, A_{t,J}=a, I_{t,J} =1\right\}\given S_t, I_{t,J} = 1\right]
\end{align*}
This completes the proof. 
\end{proof}

\subsection{Proof of Lemma \ref{lemma:asymnorm} and Lemma \ref{lemma:indasymnorm}}
\label{app:asymptotics}

We next provide a detailed proof of asymptotic normality and consistency for the C-EMEE estimator. To establish Lemma \ref{lemma:asymnorm}, we assume the following regularity conditions.

\begin{assumption}
\label{ass:consistency}
Suppose $(\alpha,\beta)\in \Theta$, where $\Theta$ is a compact subset of a Euclidean space. Suppose Equation (\ref{eq:directloglin}) has unique solution $(\hat\alpha,\hat\beta)\in \Theta$, which are consistent estimators for the solutions $(\alpha^\prime,\beta^\prime)$ that solves $E\{m_M(\alpha,\beta)\}=0$.
\end{assumption} 


Suppose Assumption \ref{consistency} holds, and $(\alpha^\prime,\beta^\prime)\in \Theta$ satisfies that $E\{m_M(\alpha^\prime,\beta^\prime)\}=0$, we have:
\begin{equation}
    \E \left[I_{t,J} e^{-A_{t,J} f_t(S_t)^\top \beta^\prime}\left(Y_{ t,\Delta,J}-e^{g_t(H_t)^\top \alpha^\prime + A_{t,J} f_t(S_t)^\top \beta^\prime}\right) W_{t,J} W_{ t,\Delta,J} \left(A_{t,J}-\tilde{p}_t(1|S_t)\right)f_t(S_t)\right] =0
\end{equation}

\begin{proof}[Proof of Consistency for direct and indirect effects]
\label{asymnorm}
For simplicity of the proof and without the loss of generality, we assume that $I_{t,J}=1$. By the law of iterated expectation:
\begin{align*}
     0=&\E \left[ e^{-A_{t,J} f_t(S_t)^\top \beta^\prime}\left(Y_{ t,\Delta,J}-e^{g_t(H_t)^\top \alpha + A_{t,J} f_t(S_t)^\top \beta^\prime}\right) W_{t,J} W_{ t,\Delta,J} \left(A_{t,J}-\tilde{p}_t(1|S_t)\right)f_t(S_t)\right] \nonumber \\
     = &\E \left(\E \left[ e^{-A_{t,J} f_t(S_t)^\top \beta^\prime}\left(Y_{ t,\Delta,J}-e^{g_t(H_t)^\top \alpha + A_{t,J} f_t(S_t)^\top \beta^\prime}\right) W_{t,J} W_{ t,\Delta,J} \left(A_{t,J}-\tilde{p}_t(1|S_t)\right)f_t(S_t)\given H_t\right]\right)\nonumber \\ 
     =&\E \left(\E \left[ e^{-A_{t,J} f_t(S_t)^\top \beta^\prime}\left(Y_{ t,\Delta,J}-e^{g_t(H_t)^\top \alpha + A_{t,J} f_t(S_t)^\top \beta^\prime}\right) W_{t,J} W_{ t,\Delta,J} \left(A_{t,J}-\tilde{p}_t(1|S_t)\right)\given H_t \right] f_t(S_t)\right)\nonumber\\ 
     =& \E \left(\E \left[ e^{- f_t(S_t)^\top \beta^\prime}\left(Y_{ t,\Delta,J}-e^{g_t(H_t)^\top \alpha + f_t(S_t)^\top \beta^\prime}\right) W_{ t,\Delta,J} \left(1-\tilde{p}_t(1|S_t)\right)\given H_t,A_{t,J}=1 \right] \tilde{p}_t(1|S_t) f_t(S_t)\right)\nonumber\\
     &- \E \left(\E \left[ \left(Y_{ t,\Delta,J}-e^{g_t(H_t)^\top \alpha^\prime }\right) W_{ t,\Delta,J} \tilde{p}_t(1|S_t)\given H_t,  A_{t,J}=0 \right]  (1-\tilde{p}_t(1|S_t)) f_t(S_t)\right)\nonumber\\
    =& \E \left[\left\{e^{- f_t(S_t)^\top \beta^\prime} \E \left(Y_{ t,\Delta,J} \given H_t,  A_{t,J}=1 \right) -\E \left(Y_{ t,\Delta,J} \given H_t,  A_{t,J}=0 \right)\right\}W_{ t,\Delta,J}  \tilde{p}_t(1|S_t)(1-\tilde{p}_t(1|S_t)) f_t(S_t)\right] \nonumber \\
    =&\E \left[e^{- f_t(S_t)^\top \beta^\prime} \E \left(Y_{ t,\Delta,J} \given H_t,  A_{t,J}=1 \right) -\E \left(Y_{ t,\Delta,J} \given H_t,  A_{t,J}=0 \right)\right]W_{ t,\Delta,J}   \tilde{p}_t(1|S_t)(1-\tilde{p}_t(1|S_t)) f_t(S_t)
\end{align*}
This indicates that 
$$
\E \left[e^{- f_t(S_t)^\top \beta^\prime} \E \left(Y_{ t,\Delta,J} \given H_t,  A_{t,J}=1 \right)\right] =\E \left[\E \left(Y_{ t,\Delta,J} \given H_t,  A_{t,J}=0 \right)\right]
$$
which is equivalent to:
$$
\E[f_t(S_t)^\top \beta^\prime]= \log \frac{\E \left(Y_{ t,\Delta,J} \given H_t,  A_{t,J}=1 \right)}{\E \left(Y_{ t,\Delta,J} \given H_t,  A_{t,J}=0 \right)}=\beta(t,H_t)
$$
Under Assumption \ref{ass:directeffect}, we have that $\beta^\prime = \beta^{\star}$. In addition, Assumption \ref{ass:consistency} implies that $\hat\beta$ converges in probability to $\beta^\star$, and this completes the proof.

We can also establish the consistency of $\hat\alpha$:

\begin{align*}
    0=&\E \left[  e^{-A_{t,J} f_t(S_t)^\top \beta^\prime}\left(Y_{ t,\Delta,J}-e^{g_t(H_t)^\top \alpha^\prime + A_{t,J} f_t(S_t)^\top \beta^\prime}\right) W_{t,J} W_{ t,\Delta,J} g_t(H_t)\right]  \\
     = &\E \left(\E \left[  e^{-A_{t,J} f_t(S_t)^\top \beta^\prime}\left(Y_{ t,\Delta,J}-e^{g_t(H_t)^\top \alpha^\prime + A_{t,J} f_t(S_t)^\top \beta^\prime}\right) W_{t,J} W_{ t,\Delta,J} g_t(H_t)\given H_t\right]\right) \\ 
     =&\E \left(\E \left[ e^{-A_{t,J} f_t(S_t)^\top \beta^\prime}\left(Y_{ t,\Delta,J}-e^{g_t(H_t)^\top \alpha^\prime + A_{t,J} f_t(S_t)^\top \beta^\prime}\right) W_{t,J} W_{ t,\Delta,J} \given H_t,  \right]  g_t(H_t)\right)\\ 
     =& \E \left[ \tilde{p}_t(1|S_t) g_t(H_t) W_{ t,\Delta,J}\left( e^{- f_t(S_t)^\top \beta^\prime} \E \left(Y_{ t,\Delta,J} \given H_t, A_{t,J}=1 \right) - e^{g_t(H_t)\alpha^\prime}\right) \right] \\
     +& \E \left[ (1-\tilde{p}_t(1|S_t)) g_t(H_t)W_{ t,\Delta,J} \left( e^{- f_t(S_t)^\top \beta^\prime} \E \left(Y_{ t,\Delta,J} \given H_t, A_{t,J}=0 \right) - e^{g_t(H_t)\alpha^\prime}\right) \right] \\
     =& \E \left[ g_t(H_t)W_{ t,\Delta,J} \left( \E \left(Y_{ t,\Delta,J} \given H_t, A_{t,J}=0 \right)- e^{g_t(H_t)\alpha^\prime}\right) \right]
\end{align*}

This indicates that:
$$
\E \left(Y_{t,\Delta,J} \given H_t,A_{t,J}=0 \right)= e^{g_t(H_t)\alpha^\prime} 
$$

Since $\exp\{g_t(H_t)^\top\alpha\}$ is a working model for $\E\{Y_{t,\Delta,J}(\Bar A_{t-1,J},0,\tilde{A}_{t+1:(t+\Delta-1),J})|H_t,A_{t,J}=0\}$, therefore $\alpha^\prime = \alpha^\star$. Then $\hat\alpha$ is a consistent estimator of $\alpha^\star$.

We next consider the indirect effect estimator. Similarly, we have the assumption:
\begin{assumption}
\label{ass:indconsistency}
Suppose $(\alpha,\beta)\in \Theta$, where $\Theta$ is a compact subset of a Euclidean space. Suppose Equation \ref{eq:indirectloglin} has unique solution $(\hat\alpha,\hat\beta)\in \Theta$, which are consistent estimators for the solutions $(\alpha^\prime,\beta^\prime)$ that solves $E\{m_M(\alpha,\beta)\}=0$.
\end{assumption} 
Suppose Assumption \ref{consistency} hold, and $(\alpha^\prime,\beta^\prime)\in \Theta$ satisfies that $E\{m_M(\alpha^\prime,\beta^\prime)\}=0$, we have:
\begin{multline}
    \E \bigg[I_{t,J} I_{t,J^\prime} W_{t,J,J^\prime}W_{t,\Delta,J,J^\prime} e^{- (1-A_{t,J})A_{t,J^\prime} f_t(S_t)^\top \beta^\prime} \\
    \left( Y_{t,\Delta,J} - e^{g_t(H_t)^\top \alpha^\prime + (1-A_{t,J})A_{t,J^\prime} f_t (S_t)^\top \beta^\prime} \right) (1-A_{t,J})(A_{t,J^\prime} - \tilde p^\star_t (1 \mid S_t) ) f_t(S_t) \bigg] =0
\end{multline}
For simplicity of the proof and without the loss of generality, we assume $I_{t,J}=1$ and $I_{t,J^\prime}=1$. Then we have the following:
\begin{align*}
    0 =&\E \Big[  W_{t,J,J^\prime}W_{t,\Delta,J,J^\prime} e^{- (1-A_{t,J})A_{t,J^\prime} f_t(S_t)^\top \beta^\prime}  \\
    & \qquad \qquad \qquad ( Y_{t,\Delta,J} - e^{g_t(H_t)^\top \alpha^\prime + (1-A_{t,J})A_{t,J^\prime} f_t (S_t)^\top \beta^\prime}) (1-A_{t,J})(A_{t,J^\prime} - \tilde p^\star_t (1 \mid S_t) ) f_t(S_t) \Big] \\
    =& \E\Big(\E \Big[  W_{t,J,J^\prime}W_{t,\Delta,J,J^\prime} e^{- (1-A_{t,J})A_{t,J^\prime} f_t(S_t)^\top \beta^\prime} \\
    & \qquad \qquad \qquad ( Y_{t,\Delta,J} - e^{g_t(H_t)^\top \alpha^\prime + (1-A_{t,J})A_{t,J^\prime} f_t (S_t)^\top \beta^\prime}) (1-A_{t,J})(A_{t,J^\prime} - \tilde p^\star_t (1 \mid S_t) )  \given H_t\Big] f_t(S_t) \Big) \\
    =& \E  \Big(\E \left[  e^{-  f_t(S_t)^\top \beta^\prime}W_{t,\Delta,J,J^\prime} ( Y_{t,\Delta,J} - e^{g_t(H_t)^\top \alpha^\prime + f_t (S_t)^\top \beta^\prime}) (1 - \tilde p^\star_t (1 \mid S_t) )  \given H_t,A_{t,J} =0, A_{t,J^\prime} = 1\right] \\
     & \qquad \qquad \qquad \qquad \qquad \qquad \qquad \qquad \qquad \qquad \qquad \tilde p_t (A_{t,J} =0, A_{t,J^\prime} = 1 \mid S_t)f_t(S_t)  \Big)  \\
    & -\E \Big(\E \left[ W_{t,\Delta,J,J^\prime}  \left( Y_{t,\Delta,J} - e^{g_t(H_t)^\top \alpha^\prime} \right) \tilde p^\star_t (1 \mid S_t)   \given H_t,A_{t,J} =0, A_{t,J^\prime}=0\right]\\
    & \qquad \qquad \qquad \qquad \qquad \qquad \qquad \qquad \qquad \qquad \qquad \tilde p_t (A_{t,J} =0, A_{t,J^\prime} = 0 \mid S_t) f_t(S_t)  \Big) \\
    =& \E \left[e^{-  f_t(S_t)^\top \beta^\prime}\E\left[Y_{t,\Delta,J}\given H_t,A_{t,J} =0, A_{t,J^\prime}=1\right] -\E\left[Y_{t,\Delta,J}\given H_t,A_{t,J} =0, A_{t,J^\prime}=0\right] \right]W_{t,\Delta,J,J^\prime} \\
    &\qquad  \big(\tilde p_t (A_{t,J} =0, A_{t,J^\prime}=0 \mid S_t)+\tilde p_t (A_{t,J} =0, A_{t,J^\prime}=1 \mid S_t)\big)\tilde p^\star_t (1 \mid S_t)(1-\tilde p^\star_t (1 \mid S_t))f_t(S_t)
\end{align*}
This indicates that 
$$
\E \left[e^{- f_t(S_t)^\top \beta^\prime} \E\left[Y_{t,\Delta,J}\given H_t,A_{t,J} =0, A_{t,J^\prime}=1\right] \right] =\E \left[\E\left[Y_{t,\Delta,J}\given H_t,A_{t,J} =0, A_{t,J^\prime}=0\right]\right]
$$
which is equivalent to:
$$
\E[f_t(S_t)^\top \beta^\prime]= \log \frac{\E\left[Y_{t,\Delta,J}\given H_t,A_{t,J} =0, A_{t,J^\prime}=1\right]}{\E\left[Y_{t,\Delta,J}\given H_t,A_{t,J} =0, A_{t,J^\prime}=0\right]}=\beta^{(IE)}(t,H_t)
$$
Under Assumption \ref{ass:indirecteffect}, we have that $\beta^\prime = \beta^{\star\star}$. In addition, Assumption \ref{ass:indconsistency} implies that $\hat\beta$ converges in probability to $\beta^{\star \star}$, and this completes the proof.

\end{proof}

\begin{proof}[Proof of Asymptotic Normality]
For the log-linear model, there is not a closed form solution; However, by Theorem 5.9 and Problem 5.27 of \cite{van2000asymptotic}. Because $m_M(\alpha,\beta)$ is continuously differentiable and hence Lipschitz continuous, Theorem 5.21 of \cite{van2000asymptotic} implies that $\sqrt{M}\{(\hat\alpha,\hat\beta)-(\alpha^\star,\beta^\star)\}$ is asymptotically normal with mean zero and covariance matrix:
$$
\E\left[\dot{m}_M(\alpha^\star,\beta^\star)\right]^{-1}\E\left[m_M(\alpha^\star,\beta^\star)m_M(\alpha^\star,\beta^\star)^\top\right]\E\left[\dot{m}_M(\alpha^\star,\beta^\star)\right]^{-1^\top}
$$

The term $Q$ depends on the derivative of the score function with respect to $\theta$:
\begin{align*}
\frac{\partial\tilde \epsilon_{t }}{\partial \alpha} &=
- e^{g_t (H_t)^\top \alpha} g_t(H_t) \\
\frac{\partial\tilde \epsilon_{t }}{\partial \beta} &=
- e^{-A_{t } f_t(S_t)^\top \beta} \times \left(Y_{t,\Delta } - e^{g_t(H_t)^\top \alpha + A_{t } f_t (S_t)^\top \beta}\right) A_{t } f_t (S_t) \\
&- e^{-A_{t } f_t(S_t)^\top \beta} \times \left( e^{g_t(H_t)^\top \alpha + A_{t } f_t (S_t)^\top \beta}\right) A_{t } f_t (S_t) \\
&= e^{-A_{t } f_t(S_t)^\top \beta} Y_{t,\Delta } A_{t } f_t(S_t) 
\end{align*}
Due to centering the matrix is block diagonal and so the covariance outer-term for $\beta$ is given by:
\begin{align*}
    \dot{m}_M(\beta) &= \sum_{t=1}^{T-\Delta +1}   e^{-A_t f_t(S_t)^\top \beta} W_t W_{t,\Delta} A_t \left(A_t-\tilde{p}_t(1|S_t)\right) \left( Y_{t,\Delta} - e^{g_t(H_t)^\top \alpha + A_{t} f_t (S_t)^\top \beta} \right)f_t(S_t)f_t(S_t)^\top +\\
    &  e^{-A_t f_t(S_t)^\top \beta} W_t W_{t,\Delta} A_t  \left(A_t-\tilde{p}_t(1|S_t)\right)e^{g_t(H_t)^\top \alpha + A_{t} f_t (S_t)^\top \beta}f_t(S_t)f_t(S_t)^\top \\
   &= \sum_{t=1}^{T-\Delta +1}   e^{-A_t f_t(S_t)^\top \beta} W_t W_{t,\Delta} A_t \left(A_t-\tilde{p}_t(1|S_t)\right) Y_{t,\Delta}f_t(S_t)f_t(S_t)^\top 
\end{align*}

Since $e^{- f_t(S_t)^\top \beta^\star} \E \left(Y_{t,\Delta} \given H_t ,A_t=1 \right) =\E \left(Y_{t,\Delta} \given H_t ,A_t=0 \right)$, thus, we have the following:\begin{align*}
    Q &= \E \left[\sum_{t=1}^{T-\Delta +1}  e^{- f_t(S_t)^\top \beta}E[Y_{t,\Delta}|H_t,A_t=1] \tilde{p}_t(1|S_t)(1-\tilde{p}_t(1|S_t))f_t(S_t)f_t(S_t)^\top   \right]\\
    &=\E \left[\sum_{t=1}^{T-\Delta +1}  E[Y_{t,\Delta}|H_t,A_t=0] \tilde{p}_t(1|S_t)(1-\tilde{p}_t(1|S_t))f_t(S_t)f_t(S_t)^\top   \right] \\
   &= \E \left[\sum_{t=1}^{T-\Delta +1} e^{g_t(H_t)^\top \alpha^\star} \tilde{p}_t(1|S_t)(1-\tilde{p}_t(1|S_t))f_t(S_t)f_t(S_t)^\top   \right]
\end{align*}

and, 
\begin{align*}
W = \E \left[   \sum_{t=1}^{T-\Delta +1} W_{t,\Delta,J} W_{t,J} \tilde \epsilon_{t,J} h_{t,J}(H_t) \times    \sum_{t=1}^{T-\Delta +1} W_{t,\Delta,J^\prime} W_{t,J^\prime} \tilde \epsilon_{t,J^\prime} h_{t,J^\prime}(H_t)^\top \right]
\end{align*}
where $\tilde \epsilon_{t,J} = e^{-A_{t,J} f_t(S_{t,J})^\top \beta} \times \left(Y_{t,\Delta,J} - e^{g_t(H_t)^\top \alpha + A_{t,J} f_t (S_{t,J})^\top \beta}\right)$, and $h_{t,J}(H_t) = ( A_{t,J} - \tilde p_t( 1 \mid S_{t,J} )) f_t (S_{t,J})$.

Similar for the indirect effect. Let $\tilde p_t (0,1 \mid S_t)$ and $\tilde p_t (0,0 \mid S_t)$ be the shorthand notations for $\tilde p_t (A_{t,J} =0, A_{t,J^\prime}=1 \mid S_t)$ and $\tilde p_t (A_{t,J} =0, A_{t,J^\prime}=0 \mid S_t)$, respectively, then the covariance outer-term for $\hat\beta^{(IE)}$ is given by:
\begin{align*}
    Q &=  \E \bigg[\sum_{t=1}^{T-\Delta +1}\big( \tilde p_t (0,1 \mid S_t)+\tilde p_t (0,0 \mid S_t)\big) \tilde p^\star_t (1 \mid S_t)(1-\tilde p^\star_t (1 \mid S_t)) e^{-f_t(S_t)^\top \beta} \\
    & \qquad \qquad \qquad \qquad \qquad \qquad \qquad \E \left[ Y_{t,\Delta,J} \mid A_{t,J} = 0, A_{t,J^\prime} = 1, H_t \right]  f_t(S_t) f_t(S_t)^\top \bigg] \\
    & =  \E \bigg[\sum_{t=1}^{T-\Delta +1} \big( \tilde p_t (0,1 \mid S_t)+\tilde p_t (0,0 \mid S_t)\big) \tilde p^\star_t (1 \mid S_t)(1-\tilde p^\star_t (1 \mid S_t))\\
    & \qquad \qquad \qquad \qquad \qquad \qquad \qquad \E \left[ Y_{t,\Delta,J} \mid A_{t,J} = 0, A_{t,J^\prime} = 0, H_t \right]  f_t(S_t) f_t(S_t)^\top \bigg] \\
    & =  \E \left[\sum_{t=1}^{T-\Delta +1} e^{g_t(H_t)^\top \alpha^\star} \big( \tilde p_t (0,1 \mid S_t)+\tilde p_t (0,0 \mid S_t)\big) \tilde p^\star_t (1 \mid S_t)(1-\tilde p^\star_t (1 \mid S_t))    f_t(S_t) f_t(S_t)^\top \right]
\end{align*}

and replace $\tilde \epsilon_{t,J,J^\prime}$ with $ e^{-A_{t,J} f_t(S_{t,J,J^\prime})^\top \beta} \left(Y_{t,\Delta,J} - e^{g_t(H_t)^\top \alpha + (1-A_{t,J})A_{t,J^\prime} f_t (S_{t,J,J^\prime})^\top \beta}\right)$, $W$ can be calculated as in the linear case in  \cite{shi2021}. This completes the proof. 
\end{proof}

\subsection{Proof of Lemma~\ref{lemma:samesies}}
\label{app:samesies}

\begin{proof}
Consider the W-matrix for the direct effect asymptotic variance:
\begin{align}
    W &= \E \left[ \frac{1}{G} \sum_{j=1}^G \sum_{t=1}^{T-\Delta +1} W_{t,j} W_{t,\Delta,j} \tilde \epsilon_{t,j} h_{t,j}(H_t) \times  \frac{1}{G} \sum_{j=1}^G \sum_{t=1}^{T-\Delta +1} W_{t,j}  W_{t,\Delta,j} \tilde\epsilon_{t,j} h_{t,j}(H_t)^\top \right] \nonumber \\
    &=  \frac{1}{G^2}\sum_{j,j^\prime} \sum_{t,t^\prime} \E\left[  W_{t,j}  W_{t,\Delta,j} \tilde\epsilon_{t,j} \left(A_{t,j}- \tilde p(1 |S_t)\right)W_{t^\prime,j^\prime} W_{t^\prime,\Delta,j^\prime}  \tilde\epsilon_{t^\prime,j^\prime} \left(A_{t^\prime,j^\prime}- \tilde p(1|S_{t^\prime})\right) f_t(S_t) f_{t^\prime}(S_{t^\prime})^\top\right]
\end{align}

where $
 \tilde\epsilon_{t,j} = e^{-A_{t,j} f_t(S_t)^\top \beta} \times \left(Y_{t,\Delta,j} - e^{g_t(H_t)^\top \alpha + A_{t,j} f_t (S_t)^\top \beta}\right)$.

Consider the cross-terms with $j \neq j^\prime$ and without loss of generality assume $t \geq t^\prime$, then
\begin{align*}
\mathbb{E} &\bigg[ \sum_{a, a^\prime}  \tilde p_t (a \mid S_t) (a - \tilde{p}_t( 1 \mid S_t)) W_{t,\Delta,j} 
\tilde p_{t^\prime} (a^\prime \mid S_{t^\prime}) (a^\prime - \tilde p_{t^\prime} (1 \mid S_{t^\prime})) W_{t^\prime,\Delta,j^\prime} \\
&\mathbb{E} \bigg[ \mathbb{E} \bigg[ \tilde\epsilon_{t,j} \tilde\epsilon_{t^\prime,j^\prime} \mid H_{t,j}, A_{t,j} = a, H_{t^\prime,j^\prime}, A_{t^\prime,j^\prime} = a^\prime \bigg] \mid S_t, S_{t^\prime} \bigg] f_t(S_t) f_{t^\prime}(S_{t^\prime})^\top
                  \bigg].
\end{align*}
Under the assumption of the error cross-term being constant in $a$ and $a^\prime$ we can re-write the above as:
\begin{align*}
&= \mathbb{E} \left[ \sum_{a, a^\prime}  \tilde p_t (a \mid S_t) (a - \tilde{p}_t( 1 \mid S_t)) W_{t,\Delta,j}
\tilde p_{t^\prime} (a^\prime \mid S_{t^\prime}) (a^\prime - \tilde p_{t^\prime} (1 \mid S_{t^\prime})) \psi (S_t, S_{t^\prime}) f_t (S_t) f_{t^\prime} (S_{t^\prime})^\top W_{t^\prime,\Delta,j^\prime} \right] \\
&= \mathbb{E} \bigg[ \psi (S_t, S_{t^\prime})W_{t,\Delta,j} W_{t^\prime,\Delta,j^\prime}f_t (S_t) f_{t^\prime} (S_{t^\prime})^\top \underbrace{\left( \sum_{a, a^\prime}  \tilde p_t (a \mid S_t) (a - \tilde{p}_t( 1 \mid S_t)) 
\tilde p_{t^\prime} (a^\prime \mid S_{t^\prime}) (a^\prime - \tilde p_{t^\prime} (1 \mid S_{t^\prime})) \right)}_{=0} \bigg] \\
&= \mathbb{E} \left[ \psi (S_t, S_{t^\prime})W_{t,\Delta,j} W_{t^\prime,\Delta,j^\prime} f_t (S_t) f_{t^\prime} (S_{t^\prime})^\top \cdot 0 \right] = 0.
\end{align*}
Therefore, we have that the $W$-matrix simplifies to
\begin{align*}
 &\mathbb{E} \bigg[ \sum_{t=1}^{T-\Delta +1} W_{t,J}W_{t,\Delta,J}\tilde \epsilon_{t,J} (A_{t,J} - \tilde{p}_t( 1 \mid S_t)) f_t(S_t) \times \sum_{t=1}^{T-\Delta +1}W_{t,J}W_{t,\Delta,J} \tilde\epsilon_{t,J} (A_{t,J} - \tilde{p}_t( 1 \mid S_t)) f_t(S_t)^\top \bigg] \\
 =&\mathbb{E} \bigg[\frac{1}{G}  \sum_{j=1}^G \bigg[ \sum_{t=1}^{T-\Delta +1} W_{t,J}W_{t,\Delta,J}\tilde \epsilon_{t,J} (A_{t,J} - \tilde{p}_t( 1 \mid S_t)) f_t(S_t) \times \\
 &~~~~~~~~~~~~~~~~~~~~~~~~\sum_{t=1}^{T-\Delta +1} W_{t,J} W_{t,\Delta,J}\tilde\epsilon_{t,J} (A_{t,J} - \tilde{p}_t( 1 \mid S_t)) f_t(S_t)^\top \bigg] \bigg] \\
 =&\mathbb{E} \bigg[ \sum_{t=1}^{T-\Delta +1} W_{t}W_{t,\Delta}\tilde \epsilon_{t} (A_{t} - \tilde{p}_t( 1 \mid S_t)) f_t(S_t) \times \sum_{t=1}^{T-\Delta +1} W_{t}W_{t,\Delta} \tilde\epsilon_{t} (A_{t} - \tilde{p}_t( 1 \mid S_t)) f_t(S_t)^\top
                  \bigg]
\end{align*}
which is the $W$ matrix as in the standard MRT analysis.  
\end{proof}

\subsection{More details on simulations}
\label{app:moresimulation}

\subsubsection{Generative model for $Z_{t,j}$ }
\label{app:markovchain}

~\

The states $Z_{t,j}$ are generated using a first-order Markov chain for each subject, with state space $\{0,1,2\}$ and transition matrix:

\begin{equation*}
    \begin{pmatrix}
0.5 & 0.25 & 0.25\\
0.25 & 0.5 & 0.25 \\
0.25 & 0.25 & 0.5
\end{pmatrix}.
\end{equation*}

This Markov chain can reach a stationary distribution, which is a uniform distribution on the state space. The true conditional treatment effect based on the generative model is:

\begin{align*}
   \beta_{\mathbf{p},1}(t;H_t) &=  \log \frac{\E \left(Y_{ t,1,j} \given H_t,  A_{t,j}=1 \right)}{\E \left(Y_{ t,1,j} \given H_t,  A_{t,j}=0 \right)}\\
   &= \frac{\left\{0.1 \mathbf{1}_{Z_{t,j}=0}+0.25 \mathbf{1}_{Z_{t,j}=1}+0.2 \mathbf{1}_{Z_{t,j}=2}  \right\}e^{(0.1+0.3Z_{t,j})}}{\left\{0.1 \mathbf{1}_{Z_{t,j}=0}+0.25 \mathbf{1}_{Z_{t,j}=1}+0.2 \mathbf{1}_{Z_{t,j}=2}  \right\}} \\
   & = 0.1+0.3Z_{t,j}.
\end{align*}

\subsubsection{Small sample size adjustment for covariance estimation}
\label{app:ssa}

~\

The robust sandwich covariance estimator~\cite{mancl2001covariance} for the entire variance matrix is given by $Q^{-1} \Lambda Q^{-1}$, and the first term,~$Q$ can be estimated by:

$$
\frac{1}{M} \sum_{m=1}^M \frac{1}{G_m}\sum_{j=1}^{G_m} D_{j,m}^\top W_{j,m} D_{j,m} 
$$
where $D_{j,m}$ is the model matrix for individual~$j$ in group $g$ associated with equation~\eqref{eq:directloglin}, and $W_{j,m}$ is a diagonal matrix of individual weights.
The middle term~$\Lambda$'s estimation is given by
$$
\frac{1}{M}\sum_{m=1}^M \left( \frac{1}{G_m}\sum_{i=1}^{G_m} D_{i,m}^\prime W_{i,m} (I_{i,m} - H_{i,m})^{-1}
e_{i,m}\right) \times \left( \frac{1}{G_m}\sum_{j=1}^{G_m} e_{j,m}^\prime (I_{j,m} - H_{j,m})^{-1} W_{j,m} D_{j,m}\right)
$$
where $I_i$ is an identity matrix of correct dimension, $e_i$ is the individual-specific residual vector and

$$
H_{j,m} = D_{j,m}
\left( \sum_{m=1}^M\frac{1}{G_m} \sum_{j=1}^{G_m} D_{j,m}^\prime W_{j,m} D_{j,m} \right)^{-1}
D_{j,m}^\prime W_{j,m}
$$
From $Q^{-1} \Lambda Q^{-1}$ we extract $\hat{\Sigma}_{\beta}$.

In the simulation study, we have all equal-size clusters, so the $G_m$ term could be extracted out of the summation. Hence, we have:

$$
\left(\sum_{m=1}^{M}\frac{1}{G_m}\sum_{j=1}^{G_m} D_{j,m}^\prime W_{j,m} D_{j,m} \right)^{-1} = G_m \left( \sum_{m=1}^M \sum_{j=1}^{G_m} D_{j,m}^\top W_{j,m} D_{j,m} \right)^{-1}
$$

\subsubsection{Indirect Effect}
\label{app:indirecteffect_factor}

~\

The total effect is defined as: $TE_{t,j} =\sum_{j\neq j^\prime}A_{t,j^\prime}\beta_{20} $, and the generative model looks like:
\begin{equation}
    \E (Y_{t,\Delta,j}\given H_t,A_{t,j}) = \left\{0.1 \mathbf{1}_{Z_{t,j}=0}+0.25 \mathbf{1}_{Z_{t,j}=1}+0.2 \mathbf{1}_{Z_{t,j}=2}  \right\}e^{A_{t,j}(0.1+0.3 \Bar{Z}_{t,g}+b_g)+TE_{t,j}}
\end{equation}

The expectation of the conditional expectation takes the form that:
\begin{align}
    \E \left[ \E \left[ Y_{t+1, J} \mid H_t, A_{t,J} = 0, A_{t,J^\prime} = 1 \right] \mid S_t \right] &= \E \left[ \mathbf{c}(z) \exp\{\beta_{20}+\sum_{i \neq {J,J^\prime}}A_{t,i}\beta_{20}\} \right] \\
    &= \mathbf{c}(z) e^{\beta_{20}} \E \left[\exp \{\sum_{i \neq {J,J^\prime}}A_{t,i}\beta_{20}\}\right]
\end{align}

Where the term $\E \left[\exp \{\sum_{i \neq {J,J^\prime}}A_{t,i}\beta_{20}\}\right]$ can be calculated as the MGF of a binomial random variable with $n=k-2$,$p$, and $t=\beta_{20}$. Therefore, this term equals to:
$$
\E \left[\exp \{\sum_{i \neq {J,J^\prime}}A_{t,i}\beta_{20}\}\right] = [pe^{\beta_{20}}+(1-p)]^{m-2}
$$

Therefore, as for the generative model, we have:
\begin{equation*}
     \E \left[ \E \left[ Y_{t+1, J} \mid H_t, A_{t,J} = 0, A_{t,J^\prime} = 1 \right] \mid S_t \right] = \left\{0.1 \mathbf{1}_{Z_{t,j}=0}+0.25 \mathbf{1}_{Z_{t,j}=1}+0.2 \mathbf{1}_{Z_{t,j}=2}  \right\} e^{\beta_{20}}[pe^{\beta_{20}}+(1-p)]^{m-2}
\end{equation*}
and
\begin{equation*}
     \E \left[ \E \left[ Y_{t+1, J} \mid H_t, A_{t,J} = 0, A_{t,J^\prime} = 0 \right] \mid S_t \right] = \left\{0.1 \mathbf{1}_{Z_{t,j}=0}+0.25 \mathbf{1}_{Z_{t,j}=1}+0.2 \mathbf{1}_{Z_{t,j}=2}  \right\} [pe^{\beta_{20}}+(1-p)]^{m-2}
\end{equation*}

And the indirect treatment effect is:

\begin{align*}
    \log  \frac{\E \left[ \E \left[ Y_{t+1, J} \mid H_t, A_{t,J} = 0, A_{t,J^\prime} = 1 \right] \mid S_t \right]}{\E \left[ \E \left[ Y_{t+1, J} \mid H_t, A_{t,J} = 0, A_{t,J^\prime} = 0 \right] \mid S_t \right]}  
&= \log  \frac{\E \left[ \exp\{\beta_{20}+\sum_{i \neq {J,J^\prime}}A_{t,i}\beta_{20}\} \mid S_t \right]}{\E \left[ \exp\{\sum_{i \neq {J,J^\prime}}A_{t,i}\beta_{20}\} \mid S_t \right]}\\
&=\log  \frac{e^{\beta_{20}}\E \left[ \exp\{\sum_{i \neq {J,J^\prime}}A_{t,i}\beta_{20}\} \mid S_t \right]}{\E \left[ \exp\{\sum_{i \neq {J,J^\prime}}A_{t,i}\beta_{20}\} \mid S_t \right]}\\
&= \beta_{20}
\end{align*}

\subsection{Simulation for Lag Treatment Effect Estimation}
\label{app:lagsimulation}
\subsubsection{Simulation setup}

~\

To evaluate the proposed estimator with $\Delta > 1$, we extend the simulation setup in Section \ref{section:sims} with $\Delta = 2$. Assuming the lag-2 treatment effect is smaller than lag-1 effect, we set $\beta_{\Delta 0} =0.05$ and $\beta_{\Delta 1} = 0.065$. Notice that the values we chose for all parameters are only intended to evaluate our proposed method, not supported by any scientific studies. Thus, the proximal response is: 
\begin{equation}
\label{eq:laggenerativemodel}
 \E (Y_{t,2,j}\given H_{t+1},A_{t+1,j}) =\left(0.1 \mathbf{1}_{Z_{t+1}=0}+0.25 \mathbf{1}_{Z_{t+1}=1}+0.2 \mathbf{1}_{Z_{t+1}=2} \right)e^{A_{t}(0.05+0.065Z_{t})+A_{t+1}(0.1+0.3Z_{t+1})}
\end{equation}


Here we identify two prespecified future (after time t) "reference" treatment regimes that define the distribution for $A_{t+1},\dots,A_{t+\Delta-1}$. The first one assigns treatment with probabilities between zero and one and corresponds to the distribution of treatments in the data we have at hand, and the second one chooses the reference regime $A_u = 1$ for $u>t$, with probability one. In this case, the lag $\Delta$ treatment effect represents the impact of a sequential treatments on the proximal response $\Delta$ time units later. Here we only present lag-2 direct causal excursion effect, the corresponding pairwise indirect effect can be constructed and estimated in the similar way.



\noindent {\bf Simulation Scenario I}. The first scenario concerns the estimation of $\beta_{\Delta}^\star$ when an important individual-level moderator exists and proximal outcomes share a random cluster-level intercept term that does not interact with treatment. We then incorporate a cluster-level random-intercept $e_g$ which follows a truncated normal distribution with $\mu =0,\sigma^2=0.5, a = -0.8, b=0.8$. However, in the data generative model, instead of directly plugging in $e_g$, we define $e_g^\prime$ to ensure that $\E[e^{e^\prime_g}]=\E[e^\mu]= 1$, in which case, $e^\prime_g = e_g - \frac{\sigma^2}{2}-\log \frac{\Phi(\frac{b}{\sigma}-\sigma)-\Phi(\frac{a}{\sigma}-\sigma)}{\Phi(\frac{b}{\sigma})-\Phi(\frac{a}{\sigma})}$. So that 
\begin{equation}
    \E (Y_{t,2,j}\given H_{t+1},A_{t+1,j}) =\left(0.1 \mathbf{1}_{Z_{t+1}=0}+0.25 \mathbf{1}_{Z_{t+1}=1}+0.2 \mathbf{1}_{Z_{t+1}=2} \right)e^{A_{t}(0.05+0.065Z_{t})+A_{t+1}(0.1+0.3Z_{t+1})+e^\prime_g}
\end{equation}
Table~\ref{tab:lagsimresults} presents the results, which shows the proposed C-EMEE approach achieves nearly unbiasedness and proper coverage under both future treatment specifications. This demonstrates that our proposed approach is applicable and stable in terms of lag treatment effect estimation with $\Delta>1$

\noindent {\bf Simulation Scenario II}. In the second scenario, we extend the above generative model to include a random cluster-level intercept term that interacts with treatment at time $t$. Similar to $e_g$ and $e_g^\prime$, $b_g$ follows a truncated normal distribution with $\mu=0, \sigma=0.5, a=-0.8, b=0.8$, and $b^\prime_g = b_g- \frac{\sigma^2}{2}-\log \frac{\Phi(\frac{b}{\sigma}-\sigma)-\Phi(\frac{a}{\sigma}-\sigma)}{\Phi(\frac{b}{\sigma})-\Phi(\frac{a}{\sigma})}$ is a random-intercept term within the treatment effect per cluster. The model takes the form of:
\begin{equation}
 \E (Y_{t,2,j}\given H_{t+1},A_{t+1,j}) =\left(0.1 \mathbf{1}_{Z_{t+1}=0}+0.25 \mathbf{1}_{Z_{t+1}=1}+0.2 \mathbf{1}_{Z_{t+1}=2} \right)e^{A_{t}(0.05+0.065Z_{t}+b^\prime_g)+A_{t+1}(0.1+0.3Z_{t+1})}
\end{equation}
Table~\ref{tab:lagsimresults} presents the results. With the presence of cluster-level random effects interact with treatment  (i.e. $b^\prime_g \not \equiv 0$), the proposed method produces nearly unbiased estimates and achieves the nominal 95\% coverage probability under both future treatment specifications. 

\noindent {\bf Simulation Scenario III}.
In the third scenario, we assume the lag treatment effect for an individual depends on the average state of all individuals in the cluster. Therefore, we define the cluster-level moderator $\bar Z_{t,g}= \frac{1}{G_g}\sum_{j=1}^{G_g}Z_{t,j}$  and consider the generative model:
\begin{equation}
 \E (Y_{t,2,j}\given H_{t+1},A_{t+1,j}) =\left(0.1 \mathbf{1}_{Z_{t+1}=0}+0.25 \mathbf{1}_{Z_{t+1}=1}+0.2 \mathbf{1}_{Z_{t+1}=2} \right)e^{A_{t}(0.05+0.065\Bar{Z}_{t,g}+b^\prime_g)+A_{t+1}(0.1+0.3Z_{t+1})}
\end{equation}
The proposed estimator again achieves the nominal 95\% coverage probability. (see Scenario III, Table~\ref{tab:lagsimresults}).

\subsubsection{Lag Treatment Effect Calculation}

~\

\noindent \textbf{Sequential Treatment Regime.}~As stated by the sequential treatment reference regime, we have the weight $W_{t,\Delta} = \frac{\pi(A_{t+1}\mid H_{t+1})}{p(A_{t+1}\mid H_{t+1})} = \frac{1[A_{t+1} = 1]}{p (A_{t+1} | H_{t+1} ) }  $. Thus, the true lag $\Delta=2$ treatment effect can simply be calculated as:
\begin{equation}
    \beta_{t,2}= \log \frac{ E \left[ Y_{t,2} \frac{1[A_{t+1} = 1]}{p (A_{t+1} | H_{t+1} ) } \mid H_{t}, A_{t} = 1 \right]}{E \left[ Y_{t,2} \frac{1[A_{t+1} = 1]}{p (A_{t+1} | H_{t+1} ) } \mid H_{t}, A_{t} = 0 \right]}
\end{equation}

Under our simulation setting, the term $E \left[ Y_{t,2} \frac{1[A_{t+1} = 1]}{p (A_{t+1} | H_{t+1} ) } \mid H_{t}, A_{t} = a \right]$ is equal to:
\begin{align*}
    E \left[ \left(0.1 \mathbf{1}_{Z_{t+1}=0}+0.25 \mathbf{1}_{Z_{t+1}=1}+0.2 \mathbf{1}_{Z_{t+1}=2}  \right)e^{A_{t}(0.05+0.065Z_{t}+b^\prime_g)+A_{t+1}(0.1+0.3Z_{t+1})} \frac{1[A_{t+1} = 1]}{p (A_{t+1} | H_{t+1} ) }\mid H_{t}, A_{t} = a \right]
\end{align*}
Splitting the expectation above to two terms, we have the following calculation:
\begin{align*}
\label{term1}
        &  E \left[ e^{A_{t}(0.05+0.065Z_{t}+b^\prime_g)} \mid H_{t}, A_{t} = a \right] \nonumber \\
        & =  e^{a(0.05+0.065Z_{t})}
\end{align*}
In the simulation setting, $A_{t+1}$ is independent with $H_t$, therefore we have the following equation:
\begin{align*}
    &E \left[ \left(0.1 \mathbf{1}_{Z_{t+1}=0}+0.25 \mathbf{1}_{Z_{t+1}=1}+0.2 \mathbf{1}_{Z_{t+1}=2}  \right) e^{A_{t+1}(0.1+0.3Z_{t+1})} \frac{1[A_{t+1} = 1]}{p (A_{t+1} | H_{t+1} ) } \mid H_{t}, A_{t} = a \right] \\
= & E \left[ \left(0.1 \mathbf{1}_{Z_{t+1}=0}+0.25 \mathbf{1}_{Z_{t+1}=1}+0.2 \mathbf{1}_{Z_{t+1}=2}  \right) e^{A_{t+1}(0.1+0.3Z_{t+1})} \frac{1[A_{t+1} = 1]}{p (A_{t+1} | H_{t+1} ) }  \right]
\end{align*}

Therefore, the true lag $\Delta=2$ treatment effect under sequential treatment regime is equal to:
\begin{align}
    \beta_{t,2}&= \log \frac{ E \left[ Y_{t,2} \frac{1[A_{t+1} = 1]}{p (A_{t+1} | H_{t+1} ) } \mid H_{t}, A_{t} = 1 \right]}{E \left[ Y_{t,2} \frac{1[A_{t+1} = 1]}{p (A_{t+1} | H_{t+1} ) } \mid H_{t}, A_{t} = 0 \right]} \nonumber \\
    & = 0.05+0.065Z_{t}
\end{align}

\noindent \textbf{Observed Distribution Treatment Regime.}~As specified by this reference treatment regime, we have future treatment reference distribution the same with the distribution of treatments in the data we have at hand, i.e., $\pi(A_{t+1}\mid H_{t+1}) = p(A_{t+1}\mid H_{t+1})$ and $W_{t,\Delta} =1$. Thus, the true lag $\Delta=2$ treatment effect can be calculated as:
\begin{equation}
    \beta^\prime_{t,2} = \log \frac{E \left[ Y_{t,2} \mid H_{t}, A_{t} = 1 \right]}{E \left[ Y_{t,2} \mid H_{t}, A_{t} = 0 \right]}
\end{equation}
Similar as above, under our simulation setting, the term $E \left[ Y_{t,2} \mid H_{t}, A_{t} = a \right]$ is equal to:
$$
    E \left[ \left(0.1 \mathbf{1}_{Z_{t+1}=0}+0.25 \mathbf{1}_{Z_{t+1}=1}+0.2 \mathbf{1}_{Z_{t+1}=2}  \right)e^{A_{t}(0.05+0.065Z_{t}+b^\prime_g)+A_{t+1}(0.1+0.3Z_{t+1})} \mid H_{t}, A_{t} = a \right] 
$$


and the true lag $\Delta=2$ treatment effect under observed treatment distribution regime is equal to $0.05+0.065Z_t$:

\subsubsection{Marginal Lag Treatment Effect Simulation Results}

~\

The choice for prespecified future reference treatment regimes is of vital importance and often time yields to different treatment effect estimations.  Following the derivations above, the fully marginal lag $\Delta=2$ treatment effect is 0.115 for both treatment reference regimes. Table~\ref{tab:lagsimresults} presents the simulation results. Overall, these results indicate that our proposed C-EMEE method produces unbiased estimation of lag-2 treatment effect, and achieves the nominal 95\% coverage probability. Closer inspection of the table shows the sequential treatment regime yields to bigger SE and RMSE estimations, which is caused by only using data points with $A_{t+1} = 1$.

\begin{table}
\centering
\caption{Simulation: C-EMEE estimators for lag $\Delta =2$ effect, under the policy of sequential treatments (ST) versus the observed treatment distribution (OTD), and comparison for Scenarios I, II, III.}
\label{tab:lagsimresults}
\begin{tabular}{cccccccc}
\hline
Scenario & Policy & \# of Clusters & Cluster Size & Bias & SE & RMSE & CP \\ \hline
\multirow{12}{*}{I} & ST & \multirow{2}{*}{25} & \multirow{2}{*}{5} & $-9.89 \times 10^{-5}$ & 0.140 & 0.128& 0.961 \\
& OTD & & &  $3.66 \times 10^{-3}$ & 0.077 & 0.075 & 0.950\\   
& ST & \multirow{2}{*}{25} & \multirow{2}{*}{10} & $6.73 \times 10^{-3}$ & 0.099 & 0.091 & 0.962 \\ 
& OTD & & &  $4.61 \times 10^{-3}$ & 0.054 & 0.055 & 0.944 \\  
& ST & \multirow{2}{*}{50} & \multirow{2}{*}{10} &$6.56 \times 10^{-3}$ & 0.067 & 0.066 & 0.955 \\ 
& OTD & & &  $3.46 \times 10^{-3}$ & 0.038 & 0.037 & 0.962\\  
& ST & \multirow{2}{*}{50} & \multirow{2}{*}{20} &$6.46 \times 10^{-3}$ & 0.048 & 0.046 & 0.956 \\ 
& OTD & & & $6.74 \times 10^{-3}$ & 0.027 & 0.028 & 0.932 \\  
& ST & \multirow{2}{*}{100} & \multirow{2}{*}{20} & $7.45 \times 10^{-3}$ & 0.033 & 0.034 & 0.955 \\ 
& OTD & & & $4.56 \times 10^{-3}$ & 0.019 & 0.020 & 0.935 \\  
& ST & \multirow{2}{*}{100} & \multirow{2}{*}{25} & $5.83 \times 10^{-3}$ & 0.030 & 0.029  & 0.948 \\
& OTD & & & $5.66 \times 10^{-3}$ & 0.017 & 0.018 & 0.941 \\ \hline
\multirow{12}{*}{II} & ST & \multirow{2}{*}{25}  & \multirow{2}{*}{5}& $-4.76 \times 10^{-3}$ & 0.164 & 0.157 & 0.956 \\ 
& OTD & & & $-1.42 \times 10^{-4}$   & 0.112 & 0.108 & 0.952 \\   
& ST & \multirow{2}{*}{25} & \multirow{2}{*}{10} & $-7.33 \times 10^{-3}$ & 0.130 & 0.123 & 0.965\\ 
& OTD & & & $-4.46 \times 10^{-3}$  & 0.096 & 0.097 & 0.942 \\  
& ST & \multirow{2}{*}{50} & \multirow{2}{*}{10} & $1.27 \times 10^{-3}$ & 0.089 & 0.089 & 0.945 \\ 
& OTD & & & $1.42 \times 10^{-3}$ & 0.068 & 0.067 & 0.946 \\  
& ST & \multirow{2}{*}{50} & \multirow{2}{*}{20} &$8.82 \times 10^{-3}$ & 0.074 & 0.073 & 0.955 \\ 
& OTD & & & $6.35 \times 10^{-3}$  & 0.062 & 0.061 & 0.951 \\  
& ST & \multirow{2}{*}{100} & \multirow{2}{*}{20} & $1.83 \times 10^{-3}$  & 0.052 & 0.051 & 0.954 \\ 
& OTD & & & $1.65 \times 10^{-3}$  & 0.044 & 0.044 & 0.955 \\  
& ST & \multirow{2}{*}{100} & \multirow{2}{*}{25} &$7.73 \times 10^{-3}$  & 0.049 & 0.050 & 0.939 \\ 
& OTD & & & $5.91 \times 10^{-3}$ & 0.042 & 0.043 & 0.940 \\ \hline
\multirow{12}{*}{III} & ST & \multirow{2}{*}{25} & \multirow{2}{*}{5} & $1.88 \times 10^{-3}$ & 0.163 & 0.154 & 0.950 \\
& OTD & & & $3.74 \times 10^{-4}$ & 0.110 & 0.111 & 0.943 \\   
& ST & \multirow{2}{*}{25} & \multirow{2}{*}{10} & $-7.76 \times 10^{-3}$ & 0.131 & 0.128 & 0.949 \\ 
& OTD & & &  $-5.93 \times 10^{-3}$ & 0.097 & 0.097 & 0.946 \\  
& ST & \multirow{2}{*}{50} & \multirow{2}{*}{10} & $2.20 \times 10^{-4}$ & 0.089 & 0.085 & 0.956 \\ 
& OTD & & & $6.58 \times 10^{-4}$ & 0.067 & 0.066 & 0.947 \\  
& ST & \multirow{2}{*}{50} & \multirow{2}{*}{20} & $-2.06 \times 10^{-3}$ & 0.074 & 0.073 & 0.948 \\ 
& OTD & & & $6.15 \times 10^{-4}$ & 0.062 & 0.062 & 0.939 \\  
& ST & \multirow{2}{*}{100} & \multirow{2}{*}{20} & $-2.25 \times 10^{-3}$ & 0.052 & 0.051 & 0.952 \\ 
& OTD & & & $-2.96 \times 10^{-3}$  & 0.043 & 0.044 & 0.949 \\  
& ST & \multirow{2}{*}{100} & \multirow{2}{*}{25} & $-2.04 \times 10^{-3}$ & 0.049 & 0.050 & 0.942 \\ 
& OTD & & & $-2.61 \times 10^{-3}$ & 0.043 & 0.043 & 0.945 \\ \hline
\end{tabular}
\end{table}

\subsection{More on Case Study}
\label{app:morecasestudy}

Following the discussion in the main paper, we consider lag $\Delta =2$ direct moderation effect analyses under both sequential treatment regime and observed treatment distribution regime. In this analysis, the same clusters are constructed as the main paper. Here, Table \ref{tab:application_lag}  presents the results. We see limited evidence of an lag-2 direct effect. 

\begin{table}
\centering
\caption{Fully marginal causal excursion effect of notifications on daily mood survey completion rate in IHS} 
\label{tab:application_lag}
\begin{tabular}{cccccccc}
\hline
Policy & Cluster  & Estimate & Std. Error  & p-value \\ \hline
\multirow{2}{*}{ST} & Institution $\times$ Specialty   & -0.011 & 0.010  & 0.30 \\ 
& Institution  & -0.005 & 0.014  & 0.73 \\  
\hline
\multirow{2}{*}{OTD} & Institution $\times$ Specialty  & -0.009  & 0.007  & 0.17\\ 
& Institution  & -0.004 & 0.009  & 0.64 \\ \hline
\end{tabular}
\end{table}

\begin{figure}
 \centerline{\includegraphics[width=6.5in]{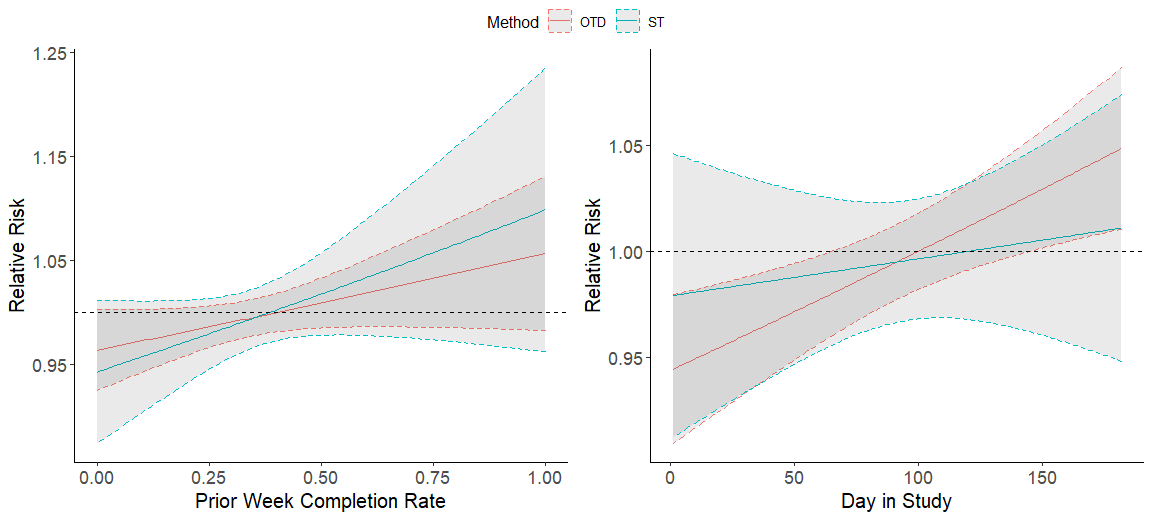}}
\caption{Moderation Analysis: \textbf{Left}: Treatment effect moderate by prior week's completion rate, and \textbf{Right}: Treatment effect moderate by day-in-study. Clusters are constructed based on a subject's membership of medical institution.}
\label{fig:two lag graphs institution}
\end{figure}

\begin{figure}
 \centerline{\includegraphics[width=6.5in]{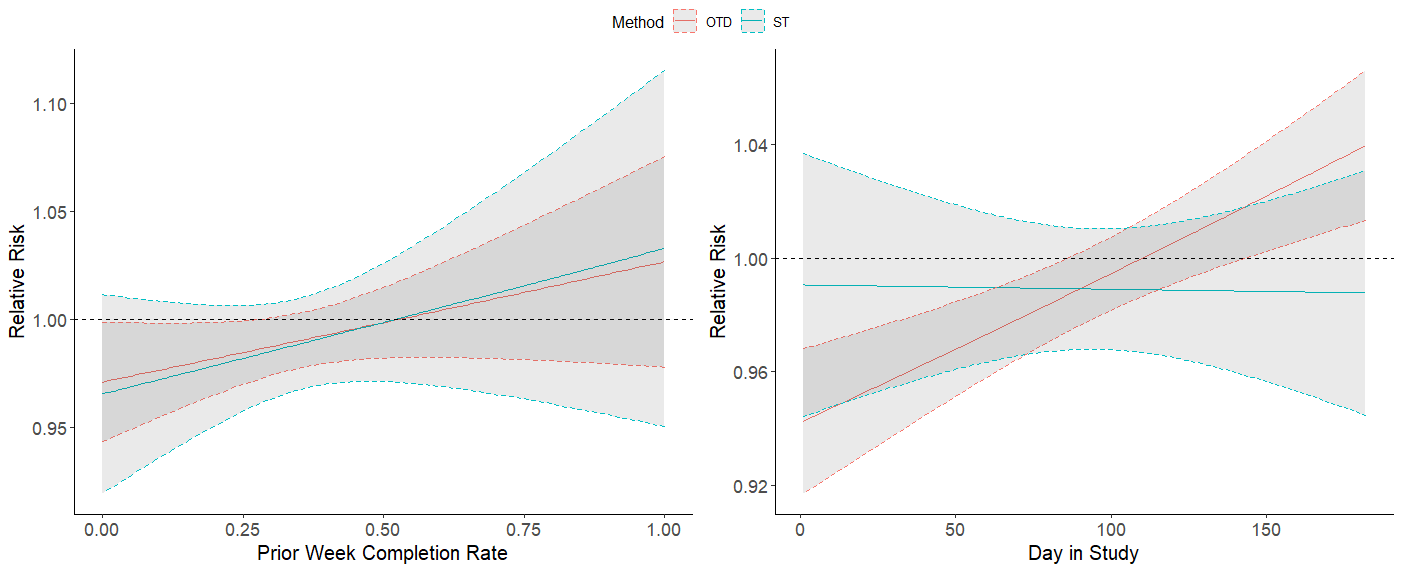}}
\caption{Moderation Analysis: \textbf{Left}: Treatment effect moderate by prior week's completion rate, and \textbf{Right}: Treatment effect moderate by day-in-study. Clusters are constructed based on a subject's joint membership of medical institution and specialty.}
\label{fig:two lag graphs institution and specialty}
\end{figure}

Figure \ref{fig:two lag graphs institution} and Figure \ref{fig:two lag graphs institution and specialty} show the lag-2 moderation effect of prior week completion rate and day in study on both clustering levels. In general, the lag-2 moderation effects are much weaker than lag-1, which indicates that the smart phone notification has a rather short time impact on users engagement. 

Combining the analyses in Section \ref{sec:casestudy}, it suggests that if the the goal is to estimate the causal excursion effect of the targeted notifications, the treatment and moderation effects from distant past contributes very minimal to the proximal outcome, therefore it is sufficient to focus on lag-1 effect only. 

\subsection{Code to Replicate Simulation and Case Study Results}
The R code used to generate the simulation experiments and case study results in this paper can be obtained at \verb"https://github.com/Herashi/binaryMRT-mHealth".

\end{document}